\documentclass[12pt]{article}
\usepackage{amsmath} 
\usepackage{multirow} 
\usepackage{amsfonts}
\usepackage{amssymb}
\def\hybrid{\topmargin -20pt    \oddsidemargin 0pt
        \headheight 0pt \headsep 0pt
        \textwidth 6.25in       
        \textheight 9.5in       
        \marginparwidth .875in
        \parskip 5pt plus 1pt   \jot = 1.5ex}
\hybrid
\numberwithin{equation}{section}
\numberwithin{table}{section}
\newcommand{\beq}{\begin{equation}}
\newcommand{\eeq}{\end{equation}}
\newcommand{\bea}{\begin{eqnarray}}
\newcommand{\eea}{\end{eqnarray}}
\newcommand{\ba}{\begin{array}}
\newcommand{\ea}{\end{array}}
\newcommand{\bt}{\begin{tabular}}
\newcommand{\et}{\end{tabular}}
\newcommand{\bc}{\begin{center}}
\newcommand{\ec}{\end{center}}

\newcommand{\Ox}{\Omega}

\newcommand{\cL}{\mathcal{L}}
\newcommand{\cK}{\mathcal{K}}

\newcommand{\cF}{\mathcal{F}}

\newcommand{\KK}{\mathcal{K}}
\newcommand{\MM}{\mathcal{M}}
\newcommand{\cM}{\mathcal M}
\newcommand{\OO}{\mathcal{O}}

\newcommand{\bi}{{\bar \imath}}
\newcommand{\ib}{{\bar\imath }}
\newcommand{\jb}{{\bar\jmath }}
\newcommand{\bj}{{\bar\jmath}}

\newcommand{\Kh}{{\hat{K}}}
\newcommand{\Lh}{{\hat{L}}}
\newcommand{\Mh}{{\hat{M}}}
\newcommand{\Nh}{{\hat{N}}}

\newcommand{\dd}{d}

\newcommand{\nn}{\nonumber}

\newcommand{\IM}{\textrm{Im} \,}
\newcommand{\RE}{\textrm{Re} \,}

\newcommand{\cref}{{\bf [check ref]}}

\newcommand{\N}{\Theta}

\newcommand{\Y}{Y}

\begin{document}

\begin{titlepage}
\begin{center}

\hfill LPTENS 04/14\\
\vskip 2cm

{\large \bf  
The effective action of $N=1$ Calabi-Yau orientifolds}

\vskip 1.5cm

{\bf Thomas W.\ Grimm$^{\rm a}$ and Jan Louis$^{\rm a,b}$}  \\
\vskip 0.3cm

${}^{\rm a}${\em II. Institut f{\"u}r Theoretische Physik\\
Universit{\"a}t Hamburg\\
Luruper Chaussee 149\\
 D-22761 Hamburg, Germany}\\

\vskip 15pt

${}^{\rm b}${\em Laboratoire de Physique Th\'eorique de l'Ecole Normale 
Sup\'erieure\\
24 rue Lhomond, \\
75231 Paris Cedex, France}\\

\vskip 10pt

 {\tt  thomas.grimm@desy.de,   jan.louis@desy.de} \\

\end{center}

\vskip 2cm

\begin{center} {\bf ABSTRACT } \end{center}

\noindent

We determine the $N=1$ low energy effective action 
for compactifications of type IIB string theory
on compact Calabi-Yau orientifolds 
in the presence of background fluxes from a Kaluza-Klein reduction.
The analysis is performed for Calabi-Yau threefolds 
which admit an isometric and holomorphic involution.
We explicitly compute the K\"ahler potential, the superpotential 
and the gauge kinetic functions and check the consistency
with $N=1$ supergravity. We find a new class of no-scale
K\"ahler potentials and show that their structure 
can be best understood in terms of a  dual formulation where
some of the chiral multiplets are replaced by linear multiplets.
For $O3$- and $O7$-planes the scalar potential is expressed
in terms of a superpotential
while for $O5$- and $O9$-planes also a $D$-term and a massive 
linear multiplet can be present.
The relation with the associated  F-theory 
compactifications is briefly discussed.

\vfill

\noindent March 2004

\end{titlepage}

\section{Introduction}
\setcounter{equation}{0}
In recent years it became clear that string theory is 
not  only a theory of fundamental strings but also contains higher-dimensional
extended objects  such as D-branes and orientifold planes \cite{JP,JPbook,AD}.
D-branes have massless excitations
associated with the attached open strings while
orientifold planes  carry no
physical degrees of freedom. 
Their relevance arises from the fact
that they can have negative tension and are often necessary ingredients
to ensure the consistency of a compactifications.
In order to cancel gravitational and electro-magnetic tadpoles
on a compact manifold in the presence of D-branes and/or
background fluxes objects with negative tension have to be included
\cite{GKP}.

Phenomenologically the most interesting case are compactifications
which lead to a (spontaneously broken) $N=1$ supersymmetry 
in four space-time dimensions ($D=4$). 
Such theories naturally arise as  Calabi-Yau compactifications
of heterotic or type~I string theories. 
In type II string compactifications 
on Calabi-Yau threefolds $Y$ one obtains instead an 
$N=2$ theory in $D=4$. However, this $N=2$ can be further broken to $N=1$
by introducing appropriate BPS D-branes and/or orientifold planes.
Turning on additional background fluxes in such compactifications
generically breaks (spontaneously) the left over $N=1$ supersymmetry
\cite{GKP}--\cite{Brustein}. 

The purpose of this paper is to determine the $N=1, D=4$ 
low energy effective action for 
Calabi-Yau orientifold compactifications in the presence of
background fluxes. To do so we use a standard Kaluza-Klein reduction
which is valid in the large volume limit.
For special classes of type IIB Calabi-Yau orientifolds
this effective action has already been determined
in refs.\ \cite{GKP,BBHL,DWG,FPo,Ferrara,BHK}.
Here we generalize the class of orientifolds
but continue to focus on type IIB string theory leaving
the discussion of type IIA and mirror symmetry to
a separate publication. For some of the cases we discuss
consistency requires the presence of background D-branes.
However, in our analysis we freeze all of the associated
massless fluctuations and solely concentrate on the couplings of the
orientifold bulk. Including D-branes fluctuations 
has recently been discussed in $N=1$ theories
in refs.\ \cite{Grana,KN,CIM,GGJL}.\footnote{
The present paper should be regarded as a companion paper
to ref.\ \cite{GGJL}.} 

More precisely,
we start from type IIB string theory and compactify on Calabi-Yau 
threefolds $Y$. In addition we 
mod out by orientation reversal of the string 
world-sheet $\Omega_p$ together with an  `internal'
symmetry $\sigma$ which acts solely on $Y$ 
but leaves the $D=4$ Minkowskian space-time untouched.
Consistency requires $\sigma$ to be  an isometric 
and holomorphic involution of $Y$ \cite{AAHV,BH}.
Hence in our analysis we focus on the class of Calabi-Yau threefolds which 
admit such an involution but within this class we leave the
threefolds arbitrary. 
One can show that for such threefolds $\sigma$ leaves
the K\"ahler form $J$ invariant but can act non-trivially on
the holomorphic three-form $\Omega$.
Depending on the transformation properties of  $\Omega$
two different symmetry operations $\mathcal{O}$
are possible \cite{Sen,DP,AAHV,BH}.
One can have either 
\beq \label{o3-projection}
\mathcal{O}_{(1)} = (-1)^{F_L} \Omega_p \, \sigma^*\ ,\qquad
\sigma^* \Omega  =  - \Omega\ ,
\eeq 
or
\beq \label{o5-projection}
\mathcal{O}_{(2)} = \Omega_p \, \sigma^*\ ,\qquad
\sigma^* \Omega  =   \Omega\ .
\eeq 
$\Omega_p$ is the world-sheet parity,
$F_L$ is the space-time fermion number
in the left-moving sector and  
$\sigma^*$ denotes the action of $\sigma$ on forms
(the  pull-back of $\sigma$).
Modding out by $\mathcal{O}_{(1)}$
leads to the possibility of having $O3$- and $O7$-planes
while modding out by $\mathcal{O}_{(2)}$ allows 
$O5$- and $O9$-planes. 

This paper is organized as follows. In order to set the stage
we briefly recall the compactification
of type IIB on Calabi-Yau threefolds in section~2 following refs.\
\cite{BGHL,Michelson,DallAgata,LM}. 
In section~3 we analyze
orientifold theories which arise when the string theory is modded out
by the symmetry (\ref{o3-projection}).
We first determine the low energy spectrum and show how it assembles
into $N=1$ supermultiplets (section~3.1). In section~3.2
we compute the effective action in the presence
of background fluxes from a Kaluza-Klein reduction and show that it
obeys the constraints of $N=1$ supergravity by explicitly determining
the K\"ahler potential, the superpotential 
and the gauge-kinetic functions.
We find that the potential is positive semi-definite 
and can be expressed entirely in terms of 
a superpotential \cite{GVW,TV,GKP,BBHL}.
The gauge kinetic functions turn out to depend holomorphically
on the complex structure deformations but they are independent 
of all other moduli. 
The K\"ahler metric is block diagonal with one factor
descending from the standard metric of the complex structure deformations \cite{CdO}.
The second  factor has a  `no-scale' form \cite{NS} and 
is a non-trivial mixing of the other moduli
including the dilaton and the K\"ahler deformations.
Aspects of this K\"ahler potential have
also been discussed recently in ref.\ \cite{DAFT} 
from a slightly different perspective.
In section 3.3 we rewrite the effective action in the linear multiplet
formalism following \cite{BGG} and show that some of its properties 
can be understood more conceptually in this dual formulation.
As a byproduct we discover a new class of no-scale K\"ahler potentials.

In section~4 we repeated the same analysis for the 
projection (\ref{o5-projection}).
In 4.1 we determine the massless spectrum while in
4.2 we compute the effective action.
As before the consistency with $N=1$ supergravity can be established by
explicitly determining the K\"ahler potential, the superpotential
and the gauge kinetic function. However, due to the different projection
(\ref{o5-projection}) the effective action turns out to have 
a different structure.
Analogously to the situation in 
$N=2$ compactifications we find that depending on the 
choice of background fluxes 
a universal two-form can become massive \cite{LM}. 
The potential is again positive semi-definite
but this time only the RR-fluxes contribute to 
the superpotential while the NS-fluxes instead lead to a $D$-term 
and a mass-term for the scalar partner of the massive two-form.
The gauge kinetic function again depends holomorphically on the 
complex structure deformations.
As before the K\"ahler potential and the couplings of
the massive two-form can be best understood in 
the linear multiplet formalism.

Section~5 contains our conclusions and some of
the details of the computations are relegated to four appendices.
In appendix~A we summarize our conventions.
In appendix~B we present the detailed computation of the scalar 
potentials while appendix~C provides the details of
the computation of the K\"ahler metric.

\section{Type IIB compactified on Calabi-Yau threefolds \label{CYsection}}

In order to set the stage for the following sections
let us briefly recall the compactification 
of type IIB supergravity on Calabi-Yau manifolds following refs.\
\cite{BGHL,Michelson,DallAgata,LM}. In sections~3 and 4 we
then repeat the analysis including the orientifold projections
(\ref{o3-projection}) and (\ref{o5-projection}), respectively.

The massless bosonic spectrum of type IIB in $D=10$ consists 
 of the dilaton 
$\hat \phi$, the metric $\hat g$ and a two-form $\hat B_2$
in the NS-NS sector and the axion $\hat l$, a two-form $\hat C_2$ and a 
four-form  $\hat C_4$ in the R-R sector.\footnote{The hats ` $\hat{}$ '
denote ten-dimensional fields.}
Using form notation (our conventions are summarized in appendix~\ref{conventions})
the type IIB low energy effective action in the $D=10$ 
Einstein frame is given by 
\cite{JPbook} 
\begin{eqnarray}\label{10d-lagr}
  S^{(10)}_{IIB}&=&
  -\int \Big(\frac{1}{2} \hat R * \mathbf{1} + \frac{1}{4} d\hat \phi\wedge *d \hat \phi
  +\frac{1}{4} e^{-\hat \phi} \hat H_3 \wedge* \hat H_3 \Big)  \\
  &-& \frac{1}{4}\int \Big( e^{2\hat \phi} d\hat l \wedge * d\hat l +
  e^{\hat \phi} \hat F_3 \wedge * \hat F_3 +
  \frac{1}{2}\hat F_{5} \wedge *\hat F_{5}\Big)
   -\frac{1}{4} \int \hat C_4 \wedge \hat H_3 \wedge \hat F_3\ ,
\nonumber  
\end{eqnarray}
where $*$ denotes the Hodge-$*$ operator and the field strengths are defined as
\begin{eqnarray}
  \hat H_3 &=& d \hat B_2\ , \qquad 
\hat F_3\ =\ d \hat C_2 - \hat l d\hat B_2\ ,
  \nonumber \\
  \hat F_5 &=& d \hat C_4 - \frac{1}{2}d\hat B_2 \wedge \hat C_2 
+ \frac{1}{2} \hat B_2
  \wedge d \hat C_2\ . \label{fieldstr}
\end{eqnarray}
The self-duality condition $\hat F_5=*\hat F_5$ is
imposed at the level of the equations of motion.

In standard Calabi-Yau compactifications one chooses the 10-dimensional
back\-ground metric to be block diagonal  or in other words 
the line element to take the form
\beq
ds^2= g_{\mu \nu} dx^{\mu} dx^{\nu}+ {g_{i\jb}} dy^i d\bar y^\jb\ ,
\label{line-metric}
\eeq
where $g_{\mu \nu}, \mu,\nu = 0,\ldots,3$ is a Minkowski metric 
and $ g_{i\jb}, i,\jb=1,\ldots,3$ is the metric on 
the Calabi-Yau manifold $Y$. 
Deformations of this metric which respect the Calabi-Yau condition
correspond to scalar fields in $D=4$. 
The deformations of the 
K\"ahler form $J = i {g}_{i\bj}\, dy^i \wedge d\bar y^\jb$ 
give rise
to $h^{(1,1)}$ real scalar fields $v^{A}(x)$ and one expands
\beq\label{transJ}
J = v^{A}(x)\, \omega_{A}  \ , \qquad A = 1, \ldots, h^{(1,1)}\ ,
\eeq 
where $\omega_A$ are harmonic $(1,1)$-forms on  $Y$
which form a basis of the cohomology group $H^{(1,1)}(Y)$.
Deformations of the complex structure are parameterized by
complex scalar fields $z^{K}(x)$
and are in one-to-one correspondence with harmonic
$(1,2)$-forms 
\beq\label{cs}
  \delta{g}_{ij} =  \frac{i}{||\Omega||^2}\, \bar z^{K} 
  (\bar \chi_{K})_{i\ib\bj}\,
  \Omega^{\ib\bj}{}_j \ , \quad K=1,\ldots,h^{(1,2)}\ ,
\eeq
where $\Omega$ is the holomorphic (3,0)-form,
$\bar\chi_{K}$ denotes a basis of $H^{(1,2)}$ and we abbreviate
$||\Omega||^2\equiv \frac1{3!}\Omega_{ijk}\bar\Omega^{ijk}$.

Using the Ansatz \eqref{line-metric}
the type IIB gauge potentials appearing in the Lagrangian 
\eqref{10d-lagr} are similarly 
expanded in terms of harmonic forms on $Y$ according to 
\begin{eqnarray}\label{CYexpansion}
  \hat B_2 &=& B_2(x) + b^A(x)\, \omega_A\ , \qquad
\hat C_2\ =\ C_2(x) + c^A (x)\,\omega_A\ , \quad A=1,\ldots,h^{(1,1)}\ , \\
  \hat C_4 &=& D_2^A(x) \wedge \omega_A + V^{\hat K}(x) \wedge
               \alpha_{\hat K} - U_{\hat K}(x) \wedge \beta^{\hat K} +
               \rho_A(x)\, \tilde \omega^A\ , 
\quad \hat K=0,\ldots,h^{(1,2)}.\nonumber
  \label{full-exp}
\end{eqnarray}
In Table \ref{CYbasis} we summarize 
the non-trivial cohomology groups on $Y$ and denote their basis elements
which appear in the expansion of the ten-dimensional fields.
As we already indicated the $\omega_A$ are harmonic $(1,1)$-forms while
the $\tilde \omega^A $ are harmonic $(2,2)$-forms which form a 
basis of $H^{(2,2)}(Y)$  dual to the $(1,1)$-forms $\omega_A$.
$(\alpha_{\hat K}, \beta^{\hat L})$ are harmonic three-forms
and form a real, symplectic basis on $H^{(3)}(Y)$
in that they satisfy
\begin{equation}\label{sbasis}
 \int  \alpha_{\hat K} \wedge \beta^{\hat L} = \delta^{\hat L}_{\hat K}\ , \qquad
 \int  \alpha_{\hat K} \wedge \alpha_{\hat L} = 0 =
 \int  \beta^{\hat K} \wedge \beta^{\hat L} \ .
\end{equation}

\begin{table}[h]
\begin{center}
\begin{tabular}{| c | c | c |} \hline
   \rule[-0.3cm]{0cm}{0.8cm} cohomology group&
   dimension & basis
   \\ \hline
   \rule[-0.3cm]{0cm}{0.8cm} $H^{(1,1)}$  &
   $h^{(1,1)}$ & $\omega_A$
   \\ \hline
   \rule[-0.3cm]{0cm}{0.8cm} $H^{(2,2)}$  & $h^{(1,1)}$ & $\tilde \omega^A$
   \\ \hline
   \rule[-0.3cm]{0cm}{0.8cm} $H^{(3)}$  & $2h^{(2,1)}+2$ &
   $(\alpha_{\hat K},\beta^{\hat L})$ \\ \hline
\rule[-0.3cm]{0cm}{0.8cm} $H^{(2,1)}$  &
   $h^{(2,1)}$ &
   $\chi_K$
   \\ \hline
\end{tabular}
\caption{\small \label{CYbasis}
\textit{Cohomology groups on $Y$ and their basis elements.}}
\end{center}
\end{table}

The four-dimensional fields appearing in the expansion \eqref{CYexpansion}
are the scalars $b^A(x)$, $c^A(x)$ and $\rho_A(x)$, 
the one-forms $V^{\hat K}(x)$ and
$U_{\hat K}(x)$ and the two-forms $B_2(x),C_2(x)$ and $D_2^A(x)$.
The self-duality condition of $\hat F_5$ eliminates half of the 
degrees of freedom in $\hat C_4$ and in this section we choose to eliminate
$D^A_2$ and $U_{\hat K}$ in favor of $\rho_A$ and $V^{\hat K}$.
Finally, the two type IIB scalars $\hat \phi, \hat l$ also appear as
scalars in $D=4$ and therefore we drop the hats henceforth
and denote them by $\phi, l$.

Altogether the massless $D=4$ spectrum consists of 
the gravity multiplet with bosonic components $(g_{\mu \nu}, V^0)$,
$h^{(2,1)}$ vector multiplets 
with bosonic components $(V^{K}, z^{K})$,
$h^{(1,1)}$ hypermultiplets with bosonic components
$(v^A, b^A, c^A, \rho_A)$
and one double-tensor multiplet \cite{BVT} with bosonic components
$(B_2, C_2, \phi, l)$ which can be dualized to an additional (universal) 
hypermultiplet. The four-dimensional spectrum is
summarized in Table \ref{tab-compIIBspec}.

\begin{table}[h]
\begin{center}
\begin{tabular}{| c | c | c |} \hline
   \rule[-0.3cm]{0cm}{0.8cm} gravity multiplet  &
   $1$ & {\small $(g_{\mu \nu},V^0)$} 
   \\ \hline
   \rule[-0.3cm]{0cm}{0.8cm} vector multiplets &
   $h^{(2,1)}$ & {\small $(V^{K}, z^{K})$}\\ \hline
   \rule[-0.3cm]{0cm}{0.8cm} hypermultiplets  &
   $h^{(1,1)}$ &
   {\small $(v^A, b^A, c^A, \rho_A)$
}\\ \hline
\rule[-0.3cm]{0cm}{0.8cm} double-tensor multiplet  &
   1 &
   {\small $(B_2, C_2,\phi,l)$ 
}\\ \hline

\end{tabular}
\caption{\small \label{tab-compIIBspec}
\textit{ $N=2$ multiplets for Type IIB supergravity compactified on a Calabi-Yau manifold.}}
\end{center}
\end{table}

The $N=2$ low energy effective action is computed by inserting
\eqref{fieldstr} -- \eqref{CYexpansion} into the action \eqref{10d-lagr}
and integrating over the Calabi-Yau manifold.
For the details we refer the reader to the literature 
\cite{BGHL,Michelson,DallAgata,LM} and only recall the results here.
One finds 
\begin{eqnarray}\label{N=2}
  S_{IIB}^{(4)} &=& \int - \frac{1}{2} R *\! {\bf 1} 
+ \frac{1}{4} \RE\cM_{\hat K \hat L} {F}^{\hat K} \wedge {F}^{\hat L} + \frac{1}{4} \IM
  \cM_{\hat K\hat L} {F}^{\hat K} \wedge * {F}^{\hat L}\nonumber\\[2mm]
&&- G_{K L} d z^K \wedge * d \bar{z}^{L} 
  - G_{AB} d v^A \wedge * d v^B
  - \frac{1}{4} d \ln \cK \wedge * d \ln \cK - \frac{1}{4} d  \phi \wedge * d 
    \phi \nonumber\\[2mm]
&& - \frac{1}{4} e^{2 \phi} \dd l \wedge * \dd l 
   - e^{-\phi} G_{AB} db^A \wedge * db^B   
  - e^{\phi} G_{AB}
  \Big(\dd c^A - l \dd b^A \Big)\wedge * \Big( \dd c^B - l \dd b^B
  \Big)\nn\\[2mm]
  && - \frac{9 G^{AD}}{4 \cK^2} \Big( \dd \rho_A - \frac{1}{2}
    \cK_{ABC} (c^B \dd b^C\!-b^B dc^C ) \Big) \wedge\! *\Big( \dd \rho_D - \frac{1}{2}
    \cK_{DEF} (c^E \dd b^F\!- b^E dc^F) \Big) \nonumber \\[2mm]
&& -\frac{\cK^2}{144}\, e^{-\phi} \dd B_2 \wedge * \dd B_2 - \frac{\cK^2}{144}\, 
   e^{\phi} 
  \Big( \dd C_2 - l \dd B_2 \Big) \wedge *\Big( \dd C_2 - l \dd B_2 \Big)
  \\[2mm]
  && +  \frac{1}{2}\Big( \dd b^A \wedge C_2 + c^A \dd B_2 \Big)\wedge
  \Big( \dd \rho_A - \cK_{ABC} c^B \dd b^C \Big) + \frac{1}{4}
  \cK_{ABC} c^A c^B dB_2 \wedge \dd b^C \nonumber \ ,
\end{eqnarray}
where $F^{\hat K}=dV^{\hat K}$. 
The gauge kinetic matrix $\cM_{\hat K\hat L}$ is related to the metric
on $H^3(Y)$ and defined by \cite{CDAF}
\bea \label{defperiod-m}
  \int \alpha_\Kh \wedge * \alpha_\Lh&=&-(\text{Im}\; \cM +(\text{Re}\; \cM)
  (\text{Im}\; \cM)^{-1}(\text{Re}\; \cM))_{\Kh \Lh}\ , \nn\\
  \int \beta^\Kh \wedge * \beta^\Lh &=&-(\text{Im}\; \cM)^{-1\ \Kh \Lh}\ ,  \\
  \int \alpha_\Kh\wedge * \beta^\Lh &=& 
  -((\text{Re}\; \cM)(\text{Im}\; \cM)^{-1})_{\Kh}^\Lh\ .  \nn
\eea
$\cM_{\hat K\hat L}$ can also be expressed in terms of the period matrix 
as \cite{dWvP,N=2review}
\begin{equation}\label{gaugeN=2}
   \cM_{\Kh \Lh}=\overline{ \mathcal{F}}_{\Kh \Lh}+2i \frac{(\text{Im}\; \mathcal{F})_{\Kh \Mh} X^\Mh
   (\text{Im}\; \mathcal{F})_{\Lh \Nh}X^\Nh }{X^\Nh(\text{Im}\; \mathcal{F})_{\Nh\Mh} 
    X^\Mh}\ ,
\end{equation} 
where $X^\Kh$ and $\mathcal{F}_\Kh$ are the periods of the holomorphic
three-form $\Omega(z)$ and 
$\mathcal{F}_{\Kh\Lh}$ is the period matrix defined as
\begin{equation}
  X^\Kh=\int_Y \Omega \wedge \beta^\Kh\ , \qquad
\mathcal{F}_\Kh= \int_Y \Omega \wedge \alpha_\Kh\ ,\qquad
\mathcal{F}_{\Kh\Lh} = \frac{\partial\mathcal{F}_\Kh}{\partial X^\Lh}\ .
  \label{def-Z}
\end{equation} 
As a consequence $\Omega$ enjoys the expansion 
$\Omega(z) =X^\Kh(z)\alpha_\Kh-\mathcal{F}_\Kh(z) \beta^\Kh$ where
both $X^\Kh(z)$ and $\mathcal{F}_\Kh(z)$ depend holomorphically on
the complex structure deformations $z^{K}$ and 
$\mathcal{F}_\Kh$ is the derivative of a holomorphic prepotential 
$\mathcal{F}$, i.e.\ 
$\mathcal{F}_\Kh = \frac{\partial\mathcal{F}}{\partial X^\Kh}$.
Finally, there
is a set of coordinates -- called special
coordinates -- where one chooses $X^\Kh = (1, z^{K})$.

The metric $G_{K L}(z,\bar z)$ which appears in \eqref{N=2}
is the metric on the space of
complex structure deformations given by \cite{CdO}
\beq\label{csmetric}
G_{KL} = 
\frac{\partial}{\partial z^K}\frac{\partial}{\partial\bar z^L}
\  K_{\rm cs}\ , \qquad
 K_{\rm cs} = -\ln\Big[- i \int_Y \Ox \wedge \bar \Ox\Big] 
= -\ln i\Big[
\bar X^\Kh\mathcal{F}_\Kh
-X^\Kh\bar{\mathcal{F}}_\Kh\Big]
\ .
\eeq
It is a special K\"ahler metric in that it is
entirely determined by the holomorphic prepotential $\mathcal{F}(z)$ 
\cite{dWvP,N=2review}.

The metric $G_{AB}$ in \eqref{N=2} is the metric 
on the space of K\"ahler deformations  defined as \cite{Strominger,CdO}
\begin{eqnarray}
 G_{AB} = \frac{3}{2\KK}
  \int_{Y}\omega_A \wedge *\omega_B = -\frac{3}{2}\left( \frac{\KK_{AB}}{\KK}-
  \frac{3}{2}\frac{\KK_A \KK_B}{\KK^2} \right)
\ ,
  \label{metric}
\end{eqnarray} 
where we abbreviated
\begin{align}\label{int-numbers}
  \KK_{ABC} &= \int_{Y}\omega_A \wedge \omega_B \wedge \omega_C\ ,& 
  \KK_{AB}  &= \int_{Y}\omega_A \wedge \omega_B \wedge J 
= \KK_{ABC}v^C\ ,&  \\
  \KK_{A}   &= \int_{Y}\omega_A \wedge J \wedge J
=\KK_{ABC}v^Bv^C \ ,
& 
  \KK &= \int_{Y}J \wedge J \wedge J
 =\KK_{ABC}v^Av^Bv^C \ .\nonumber
\end{align}
Recall that $J$ is the K\"ahler form and we repeatedly used  \eqref{transJ}
in \eqref{int-numbers}.
Note that with this convention the volume of the Calabi-Yau
manifold is given by ${\rm Vol}(Y) = \frac16\KK$.

The $D=4$ two-forms $B_2,C_2$ can be dualized to scalar fields 
such that the action \eqref{N=2} is entirely expressed in terms
of vector- and hypermultiplets. 
One finds \cite{BaggerW,dWvP,N=2review}
\begin{eqnarray}
  S_{IIB}^{(4)} & = & \int -\frac{1}{2}R * \! {\mathbf 1} 
+ \frac{1}{4} \RE\cM_{\Kh\Lh} {F}^\Kh \wedge {F}^\Lh + \frac{1}{4} \IM
  \cM_{\Kh\Lh} {F}^\Kh \wedge * {F}^\Lh\nonumber\\
&&\qquad - G_{KL} \dd z^K \wedge *\dd \bar{z}^{L} 
- h_{\hat A\hat B} \dd q^{\hat A} \wedge * \dd q^{\hat B} \ ,
  \label{action3}
\end{eqnarray}
where  $q^{\hat A}$ collectively denotes all
$h^{(1,1)}+1$ hypermultiplets and $h_{\hat A\hat B}$ is a quaternionic
metric which can be found in \cite{FS}.
In this basis the scalar manifold $\cM$ of the $N=2$ theory is
the product of a quaternionic manifold $\cM^{Q}$ spanned by the scalars
$q^{\hat A}$
in the hypermultiplets and a special
K\"ahler manifold $\cM^{SK}$ spanned by the scalars $z^K$ in the 
vector multiplets
\beq
\cM = \cM^{SK}  \times  \cM^{Q}\ .
\eeq

This ends our brief summary of type IIB compactified on
Calabi-Yau threefolds and its $N=2$ low energy effective action.
Let us now turn to the main task of this paper and impose the orientifold
projections \eqref{o3-projection}, \eqref{o5-projection}  
and derive the resulting $N=1$ 
low energy effective action.


\section{Calabi-Yau Orientifolds with $O3/O7$ planes \label{O3O7-planes}}
In this section we focus on the first type of orientifold projection 
$\mathcal{O}_{(1)} = (-1)^{F_L} \Omega_p\sigma^*$ with 
$\sigma^* \Omega  =  - \Omega$
which we already gave in \eqref{o3-projection}.
This projection has been discussed in refs.\ \cite{Sen,DP,AAHV,BH}
and here we closely follow their analysis.
$\Omega_p$ is the world sheet parity transformation 
under which the type IIB fields $\hat\phi, \hat g$ and $\hat C_2$ 
are even while
$\hat B_2,\hat l,\hat C_4$ are odd. $F_L$ is the `space-time fermion number'
in the left moving sector and therefore
$(-1)^{F_L}$  leaves the
NS-NS fields $\hat\phi, \hat g,\hat B_2 $ invariant and
changes the sign of the RR fields $\hat l,\hat C_2,\hat C_4$.
$\sigma$ is an `internal' symmetry which  acts on the 
compact Calabi-Yau manifold but leaves the $D=4$ Minkowskian
space-time invariant.
In addition, 
$\sigma$ is required to be an involution, i.e.\ to satisfy $\sigma^2={\bf 1}$,
and to act holomorphically on the Calabi-Yau coordinates \cite{AAHV,BH}.%
%
%
%
\footnote{Calabi-Yau manifolds have only 
discrete isometries. For example in the case of the quintic, 
$\sigma$ could act 
by permuting the coordinates such that
the defining equation is left invariant. 
A classification of all possible involutions
of the quintic can be found in ref.\ \cite{BH}.}
As a consequence the possible orientifold-planes in type IIB
are necessarily even-dimensional.

The induced action of $\sigma$ on forms
is denoted by the pull-back $\sigma^*$.
Since $\sigma$ is a holomorphic isometry it leaves both the metric
and the complex structure invariant. As a consequence also the K\"ahler form $J$
is invariant. However, the action of $\sigma^*$
on the holomorphic 
three-form $\Omega$ is not fixed and one can have 
$\sigma^*\Omega =\pm \Omega$.
In this section we analyze the projection \eqref{o3-projection}
where $\sigma^*\Omega =- \Omega$.

Since the four-dimensional Minkowski space is left invariant by
$\sigma$ the orientifold planes are necessarily space-time filling.
Together with the fact that they 
have to be even-dimensional (including the time direction)
this selects $O3$-, $O5$-, $O7$- or $O9$-planes as the only possibilities.
The actual
dimensionality of the orientifold plane is then determined 
by the dimensionality of the fix point set of $\sigma$ in $Y$.
In order to determine this dimensionality we need the induced
action of $\sigma$ on the tangent space at any point 
of the orientifold plane. 
Since one can always choose $\Omega \propto dy^1 \wedge dy^2 \wedge dy^3$
we see that for $\sigma^* \Omega  =  - \Omega$ 
the internal part of the orientifold plane  is 
either a point or a  surface of complex dimension two.
Together with the space-time filling part we thus can have 
$O3$- and/or $O7$-planes.
The same argument can be repeated  for $\sigma^* \Omega  =  \Omega$ 
which then leads to the possibility of 
$O5$- and/or $O9$-planes.\footnote{%
Whenever $\sigma^* = id$ 
the theory has O9-planes and coincides with type I if one introduces D9-branes
to cancel tadpoles.}

In this section we impose the projection \eqref{o3-projection}
on the type IIB theory and derive the massless spectrum 
(section~\ref{o3-spectrum}) and its 
low energy $N=1, D=4$ effective supergravity action 
(section~\ref{o3-action}).
This generalizes similar derivations already performed in refs.\
\cite{GKP,BBHL} in that we also allow for the presence of $O7$-planes.
We restrict our analysis  to the bosonic fields of the compactification
keeping in mind that the couplings of the 
fermionic partners are fixed by  supersymmetry. 
The compactification we perform is closely related to the compactification
of type IIB string theory on Calabi-Yau threefolds reviewed
in the previous section. The orientifold projection 
\eqref{o3-projection}
truncates the massless spectrum from $N=2$ to $N=1$
multiplets and also leads to a modification of the couplings
which render the low energy effective theory
compatible with $N=1$ supergravity.
Such truncation procedures from $N=2$ to $N=1$ supergravity has been carried
out from a purely supergravity point of view  
in refs.\ \cite{ADAF}.

\subsection{The massless spectrum of $N=1$ Calabi-Yau orientifolds}
\label{o3-spectrum}
Before computing the effective action let us first
determine the massless spectrum when 
the orientifold projection is taken into account and see how 
the fields assemble in $N=1$ supermultiplets \cite{BH}.
In the 
four-dimensional compactified theory
only states invariant under the projection are kept.
{}From the previous discussion one immediately infers that 
the scalars $\hat \phi,\hat l$, the metric $\hat g$ and the
four-form $\hat C_4$ are even under $(-1)^{F_L}\Omega_p$
while both two forms $\hat B_2, \hat C_2$ are odd.
Using \eqref{o3-projection} this implies that the invariant
states have to obey
\begin{equation}
\begin{array}{lcl}
\sigma^*  \hat \phi &=& \  \hat \phi\ , \\
\sigma^*   \hat g &=& \ \hat g\ , \\
\sigma^*   \hat B_2 &=& -  \hat B_2\ ,
\end{array}
\hspace{2cm}
\begin{array}{lcl}
\sigma^*   \hat l &=&  \  \hat l\ , \\
\sigma^*   \hat C_2 &=& -  \hat C_2\ ,  \\
\sigma^*  \hat C_4 &=&  \  \hat C_4\ .
\end{array}
\label{fieldtransf}
\end{equation}
In addition, $\sigma^* $ is not arbitrary but required to satisfy
\beq\label{Omegatrans}
\sigma^*\Omega =-\Omega\ .
\eeq 
Since $\sigma$ is a holomorphic involution the cohomology groups $H^{(p,q)}$
(and thus the
harmonic $(p,q)$-forms) split into two eigenspaces 
under the action of $\sigma^*$ 
\beq\label{H3split}
H^{(p,q)} = 
H^{(p,q)}_+\oplus H^{(p,q)}_-\ .
\eeq
$H^{(p,q)}_+$ has dimension $h_+^{(p,q)}$ and denotes
the even eigenspace of $\sigma^*$ while
$H^{(p,q)}_-$ has  dimension $h_-^{(p,q)}$ and denotes
the odd eigenspace of $\sigma^*$. 
The Hodge $*$-operator commutes with $\sigma^*$ since $\sigma$ preserves the
orientation and the metric of the Calabi-Yau manifold and thus the Hodge
numbers obey $h^{(1,1)}_\pm=h^{(2,2)}_\pm$. Holomorphicity of $\sigma$ 
further implies $h^{(2,1)}_\pm = h^{(1,2)}_\pm$ while
\eqref{Omegatrans} leads to 
$h^{(3,0)}_+ = h^{(0,3)}_+=0, h^{(3,0)}_- = h^{(0,3)}_-=1 $.
Furthermore, the volume-form which is proportional
to $\Omega\wedge\bar\Omega$ is invariant under $\sigma^*$ and thus one has 
$h^{(0,0)}_+ = h^{(3,3)}_+=1, h^{(0,0)}_- = h^{(3,3)}_-=0 $.
We summarize the non-trivial cohomology groups including
their basis elements in table~\ref{CYObasis}.
\begin{table}[h]
\begin{center}
\begin{tabular}{| c | c || c| c || c | c |} \hline
   \multicolumn{2}{|c||}{\rule[-0.3cm]{0cm}{0.8cm} cohomology group} &
   \multicolumn{2}{|c||}{dimension} & \multicolumn{2}{|c|}{basis}
   \\ \hline
   \rule[-0.3cm]{0cm}{0.8cm} $H^{(1,1)}_+$ & $H^{(1,1)}_-$  &
   $h^{(1,1)}_+$ & $h^{(1,1)}_- $ & $\omega_\alpha$ & $\omega_a$
   \\ \hline
   \rule[-0.3cm]{0cm}{0.8cm} $H^{(2,2)}_+$ & $H^{(2,2)}_-$  & $h^{(1,1)}_+$ & $h^{(1,1)}_-$ & 
   $\tilde \omega^\alpha$ & $\tilde \omega^a$
   \\ \hline
   \rule[-0.3cm]{0cm}{0.8cm} $H^{(2,1)}_+$  & $H^{(2,1)}_-$ 
   & $h^{(2,1)}_+$ & $h^{(2,1)}_-$ &
   $\chi_\kappa$ & $\chi_k$
   \\ \hline
   \rule[-0.3cm]{0cm}{0.8cm} $H^{(3)}_+$ & $H^{(3)}_-$  & $2h^{(2,1)}_+$ & $2h^{(2,1)}_-+2$ &
   $(\alpha_{\kappa},\beta^{\lambda})$ & $(\alpha_{\hat k},\beta^{\hat l})$ \\ \hline
\end{tabular}
\caption{\small \label{CYObasis}
\textit{Cohomology groups and their basis elements.}}
\end{center}
\end{table}

The four-dimensional invariant spectrum
is found by using the Kaluza-Klein expansion 
given in eqs.\ \eqref{transJ}, \eqref{cs} and \eqref{CYexpansion}
keeping only the fields which in addition obey \eqref{fieldtransf}.
We see immediately that both $D=4$ scalar fields arising from
$\hat\phi$ and $\hat l$ 
remain in the spectrum and as before we denote them  by 
$\phi$ and $l$.
Since $\sigma^*$  leaves 
the K\"ahler form $J$ invariant  only the
$h_+^{(1,1)}$ even K\"ahler deformations $v^\alpha$ remain in the spectrum
and we expand 
\beq\label{transJo}
J =  v^{\alpha}(x)\, \omega_{\alpha} \ ,\quad
\alpha = 1,\ldots, h_+^{(1,1)}\ ,
\eeq 
where $\omega_\alpha$ denotes a basis of $H^{(1,1)}_+$.
{} Similarly, from eq.\ \eqref{cs}  we infer that the
invariance of the metric together with
(\ref{Omegatrans}) implies that the complex structure deformations 
kept in the spectrum correspond to elements in $H^{(1,2)}_-$ 
and \eqref{cs} is replaced by
\beq\label{cso}
\delta{g}_{ij} =  \frac{i}{||\Omega||^2}\, \bar z^{k} 
(\bar \chi_{ k})_{i\ib\bj}\,
\Omega^{\ib\bj}{}_j \ , \quad k=1,\ldots,h_-^{(1,2)}\ ,
\eeq
where $\bar\chi_{k}$ denotes a basis of $H^{(1,2)}_-$.\footnote{%
In ref.\ \cite{BH} it is further shown that the
$h_-^{(1,2)}$ deformations form a smooth submanifold
of the Calabi-Yau  moduli space.}

{}From eqs.\ (\ref{fieldtransf}) we learn that in the expansion of
$\hat B_2$ and $\hat C_2$ only  odd elements survive while for 
$\hat C_4$ only even elements
are kept.
Therefore the expansion \eqref{CYexpansion} is replaced by 
\bea\label{exp1}
  \hat B_2 &=& b^a(x)\, \omega_a\ ,\qquad \hat C_2\ =\ c^a(x)\, \omega_a\ , 
\quad a=1,\ldots, h_-^{(1,1)}\ , \\
  \hat C_4 &=&  D_2^\alpha(x)\wedge \omega_\alpha
+ V^{\kappa}(x)\, \wedge \alpha_{\kappa} 
+ U_{\kappa}(x)\wedge\beta^{\kappa}+
 \rho_\alpha(x)\ \tilde \omega^\alpha\ ,\quad \kappa = 1,\ldots,h_+^{(1,2)}\ ,\nonumber
\eea
where $\omega_a$ is a basis
of $H^{(1,1)}_-$, $\tilde\omega^\alpha$ is a basis
of $H^{(2,2)}_+$ which is dual to $\omega_\alpha$, and
$(\alpha_{\kappa}, \beta^{\kappa})$ is a real, symplectic
basis of $H^{(3)}_+ = H^{(1,2)}_+ \oplus H^{(2,1)}_+$ 
(c.f.\ table~\ref{CYObasis}). 
As for Calabi-Yau compactifications
imposing the self-duality on $\hat F_5$
eliminates half of the degrees of freedom in the expansion of 
$\hat C_4$. For the one-forms  $V^{\kappa},U_\kappa$ this corresponds to
the choice of electric versus magnetic gauge potentials.
On the other hand choosing the two forms $D_2^\alpha$ 
or the scalars $\rho_\alpha$ determines
the structure of the $N=1$ multiplets to be either a linear or a chiral 
multiplet and below we discuss both cases.

Altogether the resulting $N=1$ spectrum assembles into
a gravitational
multiplet, $h_+^{(2,1)}$ vector multiplets and 
$(h_-^{(2,1)}+ h^{(1,1)}+1)$ chiral multiplets 
and is
summarized in  table~\ref{N=1spectrum} \cite{BH}.
As we already mentioned we can replace $h^{(1,1)}_+$ of the chiral multiplets
by linear multiplets.

\begin{table}[h] \label{N=1spectrum}
\begin{center}
\begin{tabular}{|c|c|c|} \hline 
 \rule[-0.3cm]{0cm}{0.8cm} 
 gravity multiplet&1&$g_{\mu \nu} $ \\ \hline
 \rule[-0.3cm]{0cm}{0.8cm} 
 vector multiplets&   $h_+^{(2,1)}$&  $V^{\kappa} $\\ \hline
 \rule[-0.3cm]{0cm}{0.8cm} 
 \multirow{3}{30mm}[-3.5mm]{chiral multiplets} &   $h_-^{(2,1)}$& $z^{k} $ \\ \cline{2-3}
 \rule[-0.3cm]{0cm}{0.8cm} 
   & 1 & $(\phi,l)$ \\\cline{2-3}
\rule[-0.3cm]{0cm}{0.8cm} 
 &  $ h^{(1,1)}_-$ &$( b^a, c^a)$ \\ 
  \hline
\rule[-0.3cm]{0cm}{0.8cm} 
{chiral/linear multiplets } &     $h^{(1,1)}_+$& $( v^\alpha, \rho_\alpha )$ \\ 
\hline
\end{tabular}
\caption{$N =1$ spectrum of $O3/O7$-orientifold compactification.}
\end{center}
\end{table} 

Compared to the $N=2$ spectrum of the Calabi-Yau compactification
given in table~\ref{tab-compIIBspec} we see that 
the graviphoton `left' the gravitational multiplet
while the $h^{(2,1)}$ $N=2$ vector multiplets decomposed
into $h_+^{(2,1)}$ $N=1$ vector multiplets plus $h_-^{(2,1)}$ 
chiral multiplets. Furthermore, the $h^{(1,1)}+1$ hypermultiplets
lost half of their physical degrees of freedom and are reduced 
into $h^{(1,1)}+1$ chiral multiplets. This is 
consistent with the theorem of \cite{AM,ADAF} where it was shown that 
any K\"ahler submanifold of a quaternionic manifold
can have at most half of its (real) dimension.

Note that the two $D=4$ two-forms $B_2$ and $C_2$ present in the $N=2$
compactification (see \eqref{CYexpansion})
have been projected out and in the expansion of  $\hat B_2$ and $\hat C_2$
only the scalar fields $c^a, b^a$ appear.
The non-vanishing of $c^a,b^a$ and $V^\kappa$ is closely related to the 
appearance of $O7$-planes. To understand this in more detail
we recall, that $O3$-planes appear 
when the fix point set of $\sigma$ is zero-dimensional in $Y$
or in other words all tangent vectors at this point are odd under
the action of $\sigma$.
This in turn implies that locally
two-forms  are even  under $\sigma^*$, while three-forms 
are odd. However, this is incompatible
with the expansions given in \eqref{exp1} for 
non-vanishing $b^a,c^a$ and $V^\kappa$. 
For a setup also including $O7$-planes we locally
get the correct transformation behavior, 
so that harmonic forms in $H^{(1,1)}_-$ and 
$H^{(2,1)}_+$ can be supported.

\subsection{The effective action in terms of chiral multiplets}\label{o3-action}
Before we derive the effective action let us recall that in type IIB
string theory it is possible 
to allow background three-form fluxes $H_3$ and $ F_3$ on the
Calabi-Yau manifold \cite{Michelson,TV,Mayr,GKP}.
The Bianchi identity together with the equation of motion imply
that $H_3$ and $ F_3$ have to be harmonic three-forms. 
In orientifold compactifications they are further constrained 
by the orientifold
projection. {}From  \eqref{fieldtransf}
we see that for the projection given in \eqref{o3-projection}
they both have to be odd under $\sigma^*$
and hence are  parameterized by elements of $H^{(3)}_-(Y)$.\footnote{This
uses the fact that the exterior derivative on $Y$ commutes with $\sigma^*$.}
It is convenient to combine the two three-forms into a complex
$G_3$  according to 
\begin{equation}\label{fluxes}
G_3 = F_3 -\tau H_3\ , \qquad \tau= l + i e^{- \phi}\ .
\end{equation}
$G_3$ can  be explicitly expanded into a symplectic basis of $H^{(3)}_-$
as 
\beq\label{G3exp}
G_3 = m^{\hat k}\alpha_{\hat k} - e_{\hat k}\beta^{\hat k}\ , \qquad
\hat k = 0,\ldots, h^{(1,2)}\ ,
\eeq
with $2(h^{(1,2)}_-+1)$ complex  flux parameters 
\beq\label{mcomplex}
m^{\hat k} = m^{\hat k}_H -\tau  m^{\hat k}_F\ , \qquad 
e_{\hat k}  = e_{\hat k}^H -\tau  e_{\hat k}^F\ .
\eeq
However, most of the time we do not need this explicit expansion and
express our results in terms of $G_3$.

The presence of background fluxes and localized sources 
requires a deviation from the standard Calabi-Yau compactifications
in that a non-trivial warp factor $e^{-2A}$ has to be included into the
Ansatz for the metric \eqref{line-metric} \cite{GKP,GP}
\beq
ds^2=e^{2A(y)} {g_{\mu \nu}}(x) dx^{\mu} dx^{\nu}+ e^{-2A(y)} 
     {g_{i\bj}}(y) dy^i d\bar y^{\bj}\ .
\label{warpmetric}
\eeq
However, in this paper we perform our analysis in the
unwarped Calabi-Yau manifold since in the large radius limit
the warp factor approaches one and the metrics 
of the two manifolds coincide  \cite{GKP,FP}. 
This in turn also implies that the metrics on the moduli space
of deformations agree and as a consequence the kinetic terms
in the low
energy effective actions are the same. The difference appears 
in the potential when some of the 
Calabi-Yau zero modes are rendered massive. 

The four-dimensional effective action is computed 
by redoing the Kaluza-Klein reduction of the ten-dimensional
type IIB action given in \eqref{10d-lagr} for the truncated orientifold
spectrum including the background fluxes \eqref{fluxes}.
Inserting \eqref{exp1} into \eqref{fieldstr}
we arrive at
\begin{eqnarray} \label{HFF}
  \hat H_3 &=& db^a \wedge \omega_a + H_3\ , \qquad
  \hat F_3 = dc^a \wedge \omega_a - l\, 
  db^a \wedge \omega_a + F_3 - l\, H_3 \ , \\
  \hat F_5 &=& dD_2^\alpha \wedge \omega_\alpha + 
  dV^{\kappa} \wedge \alpha_{\kappa} 
  - dU_{\kappa} \wedge \beta^{\kappa} + 
  d\rho_\alpha \tilde \omega^\alpha-\frac{1}{2}(c^a db^b - b^a dc^b)\wedge \omega_a 
  \wedge \omega_b \ , \nonumber
\end{eqnarray}
where we   allowed for the presence of background fluxes  
$H_3$ and $ F_3$.\footnote{Note that $H_3$ and $F_3$ do not effect 
$\hat F_5$ since the only possible terms would be of
the form $H_3 \wedge C_2 $ or  $B_2 \wedge F_3$ 
but both $C_2$ and $B_2$ are projected 
out by the orientifold projection.} 
The next step is to insert \eqref{HFF} into \eqref{10d-lagr}
and perform the integration over $Y$. In order to do so we first need to 
reconsider the structure of the metrics \eqref{csmetric},
\eqref{metric} and the intersection numbers \eqref{int-numbers}
for the orientifold.

Let us start with the complex structure deformations.
Due to the split of the cohomology
$H^{(3)}= H^{(3)}_+\oplus H^{(3)}_-$ the real symplectic basis 
$(\alpha_\Kh , \beta^\Lh)$
of $H^{(3)}$ also splits into $(\alpha_{\kappa} , \beta^{\lambda})$
of $H^{(3)}_+$ and $(\alpha_{\hat k} , \beta^{\hat l})$
of $H^{(3)}_-$. Eqs.\ \eqref{sbasis} continue to hold which implies
that both basis are symplectic and obey
\begin{equation}\label{sbasiso}
 \int \alpha_{\kappa} \wedge \beta^{\lambda} 
= \delta^{\lambda}_{\kappa}\ ,
\qquad
\int \alpha_{\hat k} \wedge \beta^{\hat l} 
= \delta^{\hat l}_{\hat k}\ ,
\end{equation}
with all other intersections vanishing.
Furthermore,
we saw in the previous section that out of $h^{(2,1)}$ 
complex structure deformation $z^{K}$ only $h^{(2,1)}_-$ 
(denoted by $z^{k}$) survived. 
The three-form $\Omega$ being an element of $H^{(3)}_-$ can thus be expanded
according to 
\begin{equation}
\Omega(z^k) = X^{\hat k}\alpha_{\hat k} - \mathcal{F}_{\hat k}
\beta^{\hat k}\ ,  \qquad \hat k = 0,\ldots,h^{(1,2)}_-\ ,
\label{cond-1}
\end{equation} 
while the `other' periods $(X^{\kappa},\mathcal{F}_{\kappa})$
vanish
\beq
 X^{\kappa}= \int_Y \Omega\wedge\beta^\kappa = 0 \ , \qquad
\mathcal{F}_{\kappa}\big|_{z^{\kappa}=0} 
=  \int_Y \Omega\wedge\alpha_\kappa = 0 \ , \qquad
\kappa = 1,\ldots,h^{(1,2)}_+\ .
\eeq
As a consequence the metric on the space 
of complex structure deformations reduces to 
\beq\label{csmetrico}
G_{ kl} = 
\frac{\partial}{\partial z^{k}}
\frac{\partial}{\partial\bar z^{l}}
\  K_{\rm cs}\ , \qquad
 K_{\rm cs} = -\ln\Big[ - i \int_Y \Ox \wedge \bar \Ox\Big] 
= -\ln i\Big[\bar X^{\hat k}\mathcal{F}_{\hat k}    - X^{\hat k}\bar{\mathcal{F}}_{\hat k} \Big]
\ ,
\eeq
replacing \eqref{csmetric}.

Let us now turn to the K\"ahler deformations.
Corresponding to the decomposition  
$H^{(1,1)}=H^{(1,1)}_+\oplus H^{(1,1)}_-$ also
the harmonic (1,1)-forms $\omega_A$
split into 
$\omega_A = (\omega_\alpha, \omega_a)$
such that 
$\omega_\alpha$ is a basis of 
$H^{(1,1)}_+$ and 
$\omega_a$ is a basis of 
$H^{(1,1)}_-$. This in turn results in a decomposition of the intersection 
numbers $\KK_{ABC}$ given in \eqref{int-numbers}.
Under the orientifold projection
only $\KK_{\alpha\beta\gamma}$ and $\KK_{\alpha bc}$ can be non-zero
while $\KK_{\alpha \beta c}= \KK_{abc} =0$ has to hold. 
Since the K\"ahler-form $J$ is invariant 
we also conclude from \eqref{int-numbers}
that $\KK_{\alpha b}=0=\KK_{a}$. To summarize,
keeping only the invariant intersection numbers results in
\begin{eqnarray}\label{constr}
  \KK_{\alpha \beta c}= \KK_{abc} =\KK_{\alpha b}=\KK_{a}=0\ ,
\end{eqnarray}
while all the other intersection numbers can be non-vanishing.\footnote{From
a supergravity point of view this 
has been also observed in refs.\ \cite{ADAF}.}
Inserting \eqref{constr} into \eqref{metric} we derive
\begin{eqnarray} \label{splitmetr}
  G_{\alpha \beta}=
  -\frac{3}{2}\left( \frac{\KK_{\alpha \beta}}{\KK}-
  \frac{3}{2}\frac{\KK_\alpha \KK_\beta}{\KK^2} \right)\ , \qquad
  G_{a b}=-\frac{3}{2} \frac{\KK_{a b}}{\KK}\ , \qquad
G_{\alpha b}\ =\ G_{a \beta}\ =\ 0\ ,
\end{eqnarray}
where
\begin{equation}\label{intO3}
  \KK_{\alpha\beta}=\KK_{\alpha\beta\gamma}\; v^\gamma\ , 
\quad \ 
\KK_{ab}=\KK_{ab\gamma}\; v^\gamma\ ,\quad
\KK_{\alpha}=\KK_{\alpha \beta\gamma}\; v^\beta v^\gamma\ ,
  \quad \KK=\KK_{\alpha \beta \gamma} \; v^\alpha v^\beta v^\gamma
\ .
\end{equation}
We see that the metric $G_{AB}$ given in \eqref{metric}
is block-diagonal with respect to the 
decomposition $H^{(1,1)}=H^{(1,1)}_+\oplus H^{(1,1)}_-$.
For later use let us also record the inverse metrics
\begin{eqnarray}\label{Ginvers}
  G^{\alpha \beta}
  =  -\frac{2}{3} \KK \KK^{\alpha \beta} + 2 v^\alpha v^\beta\ ,
\qquad
  G^{a b} =
   - \frac{2}{3} \KK \KK^{a b}\ ,
\end{eqnarray}
where $\KK^{\alpha \beta}$ and $\KK^{a b}$ are the inverse 
of $\KK_{\alpha \beta}$ and
$\KK_{a b}$, respectively.

To calculate the four-dimensional action we  insert 
(\ref{transJo}), (\ref{cso}) and (\ref{HFF}) into (\ref{10d-lagr}), 
integrate over $Y$ using 
(\ref{csmetrico})--(\ref{Ginvers}).  
Furthermore we impose the self-duality condition 
$\hat F_5 = * \hat F_5$
by adding the following total derivative to the action \cite{DallAgata}
\begin{equation}
  \delta S^{(4)}_{O3/O7} = \frac{1}{4} dV^{\kappa} \wedge dU_{\kappa} + 
  \frac{1}{4} dD_2^\alpha \wedge d\rho_\alpha\ .
  \label{totalderiv} 
\end{equation} 
Then the  equation of motions for 
$D_2^\alpha$ and $U_{\kappa}$ 
(or equivalently for $\rho_\alpha,V^\kappa$)
coincide with the self-duality
condition and we can consistently eliminate $D_2^\alpha$ and $U_{\kappa}$
(or $\rho_\alpha,V^\kappa$)
by inserting their equations of motions into the action \cite{DallAgata}.
Keeping $V^\kappa$ corresponds to the choice
of expressing the action in terms of an electric instead 
of a magnetic gauge potential $U_\kappa$. 
Choosing to eliminate $D_2^\alpha$ or 
$\rho_\alpha$ corresponds to the choice of expressing the action 
in terms of linear or chiral multiplets. The standard
$N=1$ supergravity formulation \cite{WB,GGRS} uses chiral multiplets 
and thus from this point of view 
it is more convenient to eliminate $D_2^\alpha$ in favor of 
$\rho_\alpha$ and express everything in terms of chiral multiplets. However,
the resulting geometry of the $N=1$ moduli space
can be understood more conceptually by using linear multiplets.
For this reason we supplement our analysis with a discussion of the
low energy effective action in terms of
linear multiplets in section~\ref{Linearm}.

Eliminating $D_2^\alpha$ and $U_{\kappa}$ by its equations of motion
and  performing a Weyl rescaling of the 
four-dimensional metric 
$g_{\mu \nu} \rightarrow \frac{\KK}{6}g_{\mu \nu}$ to 
obtain the canonically normalized Einstein-Hilbert term
we arrive at\footnote{For vanishing fluxes this action
can also be obtained by  directly inserting the truncated spectrum 
into \eqref{N=2} and the fluxes only add the potential $V$.} 
\begin{eqnarray}\label{S_scalar}
S^{(4)}_{O3/O7} &=&\int_{\mathbb{M}_{3,1}} -\frac{1}{2}R*\mathbf{1}-
  G_{k { l}} \; dz^{k} \wedge *d\bar z^{l}
  -G_{\alpha \beta} \; dv^\alpha \wedge *dv^\beta 
  - \frac{1}{4}d\, \text{ln} \KK \wedge * d\, \text{ln} \KK \nonumber \\
  &&-\frac{1}{4} d\phi \wedge * d\phi
    -\frac{1}{4}e^{2 \phi} dl \wedge * dl 
  -e^{-\phi } G_{ab}\; db^a \wedge * db^b\nonumber \\
  && -   e^{ \phi}
  G_{ab}\left(dc^a-l db^a \right) \wedge *\left(dc^b-l db^b \right)\\
  &&-\frac{9G^{\alpha \beta}}{4\KK^2}  \Big(d\rho_\alpha-
  \frac{1}{2}\KK_{\alpha a b}(c^a db^b-b^a dc^b ) \Big)
  \wedge
  *\Big(d\rho_\beta-\frac{1}{2}\KK_{\beta cd}(c^c db^d-b^c dc^d ) 
\Big)\nonumber\\
 & &+\frac{1}{4}\text{Im}\; \cM_{\kappa \lambda}\; 
    F^{\kappa}\wedge *F^{\lambda}
     +\frac{1}{4}\text{Re}\; \cM_{\kappa \lambda}\;
     F^{\kappa}\wedge F^{\lambda} - V*\mathbf{1}\ , \nonumber
\end{eqnarray}
where $F^\kappa = dV^\kappa$ and 
${\cM}_{\kappa \lambda}$ is the 
$N=2$ gauge kinetic matrix given in \eqref{gaugeN=2}
evaluated at ${z^{\kappa}=\bar z^{\kappa}}=0$.
The potential $V$ is manifestly positive semi-definite and found to be
\cite{TV,GKP,BBHL,DWG}
\beq\label{potential37}
 V= 
  \frac{18i\ e^{\phi}}
  {\KK^2 \int \Omega \wedge \bar \Omega }
  \left( \int \Omega \wedge \bar G_3 \int \bar \Omega \wedge G_3 
  + G^{k { l}} \int \chi_{ k} \wedge G_3 
  \int \bar \chi_{ l} \wedge \bar G_3 \right)\ ,
\eeq
where
$\chi_{k}$ is a basis of $H^{(2,1)}_-$ as defined in \eqref{cso} and 
the background flux $G_3$ is defined in \eqref{fluxes}.
For completeness we include the details of the computation
of $V$ in appendix~\ref{scalarpot}

Strictly speaking the additional term 
$\cL^{(4)}_{\text{top}} \sim 
  \int_{Y}H_3 \wedge F_3 $
arises in the Kaluza-Klein reduction. 
However, consistency of the compactifications requires its cancellation 
against Wess-Zumino like couplings of the orientifold planes 
to the R-R flux \cite{GKP}. 

Our next task is to transform the action \eqref{S_scalar} into the standard
$N=1$ supergravity form where it is 
expressed in terms of a K\"ahler potential $K$, 
a holomorphic superpotential $W$ and the holomorphic gauge-kinetic coupling 
functions $f$ as follows \cite{WB,GGRS}
\beq\label{N=1action}
  S^{(4)} = -\int \tfrac{1}{2}R * \mathbf{1} +
  K_{I \bar J} DM^I \wedge * D\bar M^{\bar J}  
  + \tfrac{1}{2}\text{Re}f_{\kappa \lambda}\ 
  F^{\kappa} \wedge * F^{\lambda}  
  + \tfrac{1}{2}\text{Im} f_{\kappa \lambda}\ 
  F^{\kappa} \wedge F^{\lambda} + V*\mathbf{1}\ ,
\eeq
where
\beq\label{N=1pot}
V=
e^K \big( K^{I\bar J} D_I W {D_{\bar J} \bar W}-3|W|^2 \big)
+\tfrac{1}{2}\, 
(\text{Re}\; f)^{-1\ \kappa\lambda} D_{\kappa} D_{\lambda}
\ .
\eeq
Here the $M^I$ collectively denote  all
complex scalars in the theory  and 
$K_{I \bar J}$ is a K\"ahler metric satisfying
$  K_{I\bar J} = \partial_I \bar\partial_{\bar J} K(M,\bar M)$.
The scalar potential is expressed in terms of the 
K\"ahler-covariant derivative $D_I W= \partial_I W + 
(\partial_I K) W$. 

We first need to find a complex structure on the space of
scalar fields such that the metric computed in \eqref{S_scalar}
is manifestly K\"ahler. 
As we saw in \eqref{csmetrico} the 
complex structure deformations $z^{k}$ are already good K\"ahler
coordinates with $G_{kl}$ being the appropriate K\"ahler metric. 
For the remaining fields the definition of 
the K\"ahler coordinates is not so obvious. 
Guided by refs.\ \cite{HL,BBHL} we define
\bea \label{tau}
  \tau &=&l+i e^{-\phi}\ , \qquad
  G^a = c^a -\tau b^a\ , \nonumber\\
  T_\alpha &=& \frac{3i}{2}\, \rho_\alpha 
  + \frac{3}{4}\KK_{\alpha}(v)  - \frac{3}{2}\, \zeta_\alpha(\tau,\bar\tau,G,\bar G) \ ,
\eea
where\footnote{The definition of $\zeta_\alpha$ is unique up to a constant
which does not enter into the metric. The possibility of a non-zero constant
is important for the formulation in terms of linear multiplets in 
section~\ref{Linearm}.}  
\beq\label{zetadef}
\KK_{\alpha} =  \KK_{\alpha\beta\gamma} v^\beta v^\gamma \ ,\qquad
\zeta_\alpha = - \frac{i}{2(\tau-\bar \tau)}\ \KK_{\alpha b c}G^b (G- \bar G)^c\ .
\eeq
In appendix~\ref{Kmetrics3} we check explicitly 
that in terms of these coordinates
the metric of \eqref{S_scalar} is K\"ahler with the K\"ahler potential
\begin{eqnarray} \label{kaehlerpot-O7-1}
   K&=&K_{\rm cs}(z,\bar z)   
   + K_{k}(\tau,T,G)\ , \quad K_{\rm cs}\ =\ 
    -\text{ln}\Big[-i\int\Omega(z) \wedge \bar \Omega(\bar z) \Big] \ ,\\
  K_{k}\ &=& \ -\text{ln}\big[-i(\tau - \bar \tau)\big]
  - 2 \text{ln}\Big[\tfrac16\cK(\tau,T,G)\Big]\ .\nonumber
\end{eqnarray}
$\cK \equiv\KK_{\alpha\beta\gamma} v^\alpha v^\beta v^\gamma 
\equiv 6\, \text{Vol}(\Y)$ 
should be understood 
as a function of the K\"ahler coordinates $(\tau,T,G)$ 
which enter by solving (\ref{tau}) for $v^\alpha$ in terms of $(\tau,T,G)$.
Unfortunately this solution cannot be given explicitly and therefore $\cK$ is known
only implicitly via $v^\alpha(\tau,T,G)$.\footnote{This is in complete analogy
to the situation encountered in compactifications
of M-theory on Calabi-Yau fourfolds studied in \cite{HL}. 
This is no coincidence and can be understood from the fact
that this theory can be lifted to F-theory on Calabi-Yau fourfolds
which in a specific limit  is related to  orientifold
compactifications of type IIB \cite{Sen}.} 
%
In the  next section we show that the definition of the
K\"ahler coordinates \eqref{tau} and the K\"ahler potential
\eqref{kaehlerpot-O7-1} can be understood somewhat
more conceptually in a dual formalism using linear multiplets 
$L^\alpha$ instead of the chiral multiplets $T_\alpha$.
From a slightly different perspective these `dual'
 variables have recently also been discussed in ref.\ \cite{DAFT}.

Let us return to the K\"ahler potential (\ref{kaehlerpot-O7-1}).
The first two terms are the standard
K\"ahler potentials for the complex structure deformations
and the dilaton, respectively. 
$\cK$ also depends on $\tau$ and therefore the metric mixes $\tau$
with $T_\alpha$ and $G^a$. It is block diagonal in the 
complex structure deformations which do not mix with the other scalars.
Thus, the moduli space has the form
\beq \label{modulispace}
\cM = \cM^{h_-^{(1,2)}}_{\rm cs}\times\, \cM^{h^{(1,1)_{\vphantom{+}}}+1}_{\rm k}\ ,
\eeq
where each factor is a K\"ahler manifold and 
$\cM^{h_-^{(1,2)}}_{\rm cs}$ even is a special K\"ahler manifold
in that $K_{\rm cs}$ satisfies \eqref{csmetrico}.

Although not immediately obvious from its definition
$K_{k}$ obeys a no-scale type condition in that it satisfies 
\beq\label{NScond}
    \frac{\partial K_{k}}{\partial M^I}\,  (K_{k}^{-1})^{ I\bar J}\, 
    \frac{\partial K_{k}}{\partial \bar M^{\bar J}} = 4\ ,
\eeq
where $M^I = (\tau,G^a,T_\alpha,z^k)$. For $G^a=0$
this has already been observed in \cite{GKP,BBHL,DWG,DAFT} while 
for $G^a\neq0$ we explicitly check 
\eqref{NScond} in appendix \ref{Kmetrics3}. From \eqref{N=1pot}  we see that 
\eqref{NScond} implies $V\ge 0$ which we also show 
in the linear multiplet formalism in the next section \ref{Linearm}.
For $\tau=\text{const.}$ the right hand side of \eqref{NScond} is found to
be equal to $3$ as it is the case in the standard no-scale K\"ahler potentials
of \cite{NS}.\footnote{We like to thank Geert Smet and Joris Van Den Bergh for drawing our
attention to an error in an earlier version of this paper.}

Let us relate \eqref{kaehlerpot-O7-1} to the known K\"ahler potentials 
in the literature. 
First of all, for $G^a=0$ and thus 
$T_\alpha=\frac{3i}{2} \rho_\alpha + \frac{3}{4}\KK_\alpha $
 the K\"ahler potential 
\eqref{kaehlerpot-O7-1} reduce to the one given in \cite{BBHL}.\footnote{%
Recall that this corresponds to the situation where only 
$O3$-planes are present.}
Secondly, for one overall K\"ahler modulus $v$ parameterizing the volume
(i.e. for $h^{(1,1)}_+=1$, $T_{\alpha}\equiv T$), but keeping 
all $h^{(1,1)}_-$ moduli, eq.\ \eqref{tau} can be solved for
$v$ and one finds
\beq\label{KoneT}
  -2 \ln\; \cK = -3 \ln \frac{2}{3}\left[ T + \bar T
  - \frac{3i}{4(\tau - \bar \tau)} \cK_{1 a b} (G-\bar G)^a (G-\bar G)^b \right]\ .
\eeq
If in addition we set  $G^a=0$ and defines 
$-\frac{3i}{2}\tilde \rho \equiv T = 
\frac{3i}{2} \rho +\frac{3}{4} \KK^{2/3}$ 
\eqref{KoneT} reduce to 
$K=-3 \text{ln}(-i(\tilde \rho - \overline{ \tilde \rho} ))$
which coincides with the K\"ahler potential 
determined in \cite{GKP}.\footnote{Since 
$\KK_{\alpha \beta \gamma} v^\alpha v^\beta v^\gamma=\KK$ we
find for the case of only one K\"ahler modulus $v=\KK^{1/3}$.}

Before we turn to the discussion of the gauge kinetic functions
and the superpotential let us note that $K$ is invariant under
the $SL(2,{\bf R})$ transformations inherited from the
ten-dimensional type IIB theory. In the orientifold theory
this symmetry acts on $\tau$ by fractional linear transformations 
$\tau \to \frac{a\tau + b}{c\tau +d}$
exactly as in $D=10$ and transforms $(b^a,c^a)$ as a doublet.
Under the $SL(2,{\bf R})$ only the second term of $K$ 
given in \eqref{kaehlerpot-O7-1}
transforms but this transformation is just a K\"ahler transformation.
The two other terms are invariant as can be seen from \eqref{tau}
and the fact that $v^\alpha$ and $z^k$  are invariant.\footnote{%
Contrary to the Calabi-Yau compactifications discussed in 
section~\ref{CYsection} the K\"ahler moduli $v^\alpha$ 
are not redefined with powers of the dilaton \cite{BGHL} and
as a consequence they stay $SL(2,{\bf R})$ invariant.}

Our next task is to determine
the gauge-kinetic coupling functions $f_{\kappa \lambda}$ 
and show that they
are holomorphic in the moduli.  By comparing the actions
\eqref{S_scalar} and \eqref{N=1action} we find
\begin{eqnarray}
  f_{\kappa \lambda}
=-\frac{i}{2} \,
\bar{\cM}_{\kappa \lambda}\Big|_{z^{\kappa}=0=\bar z^{\kappa}}\ ,
\end{eqnarray} 
where 
${\cM}_{\kappa \lambda}$ is the 
$N=2$ gauge kinetic matrix given in \eqref{gaugeN=2}
evaluated at ${z^{\kappa}=\bar z^{\kappa}}=0$.
Its holomorphicity in the complex structure deformations $z^k$ is not
immediately obvious but can be shown by using
\eqref{defperiod-m} and  \eqref{gaugeN=2}.
More precisely, from \eqref{defperiod-m} together with 
the decomposition of $H^{(3)}$ expressed by \eqref{H3split}
and \eqref{sbasiso} we infer that ${\cM}_{\hat K \hat L}$
is block diagonal or in other words
${\cM}_{\kappa \hat l} = 0.$ Multiplying ${\cM}_{\kappa \hat l}$
with $X^{\hat l}$ and using $X^\lambda=0$ together with
\eqref{gaugeN=2} we further conclude 
\beq\label{Fdiag}
{\cF}_{\kappa \hat l}\Big|_{z^{\hat\kappa}=0=\bar z^{\hat\kappa}}=0\ .
\eeq
Finally inserting \eqref{cond-1} and \eqref{Fdiag}
into \eqref{gaugeN=2}
we arrive at 
\begin{eqnarray}\label{fholo}
  f_{\kappa \lambda} (z^k)
=-\frac{1}{2} i
\mathcal{F}_{\kappa \lambda}\Big|_{z^{\kappa}=0=\bar z^{\kappa}}\ ,
\end{eqnarray} 
which is manifestly holomorphic since $\mathcal{F}_{\kappa \lambda}(z^k)$
are holomorphic functions of the complex structure
deformations $z^k$.

Finally, we show that the potential \eqref{potential37}
can be derived from a superpotential $W$ via the expression
given in \eqref{N=1pot} with vanishing $D$-term $D_\kappa=0$. 
For orientifolds with $c^a=b^a=0$ (corresponding to  only $O3$-planes present)
$W$  was  shown to be \cite{GVW,TV,GKP,BBHL,DWG}
\begin{equation}
  W(\tau,z^k) =  \int_{Y} \Omega \wedge G_3\ .
  \label{superpot}
\end{equation}
This continues to be the correct superpotential
also if  $c^a$ and $b^a$ are in the spectrum.
Indeed, using \eqref{tau}, \eqref{kaehlerpot-O7-1}, (\ref{def-chi})
and \eqref{K-deriv} one computes the
K\"ahler-covariant derivatives  to be 
\begin{eqnarray}
  D_{\tau}W&=& \frac{i}{2}e^{\phi} \int \Omega \wedge \bar G_3
             +i G_{ab}b^a b^b\ W\ , \qquad
  D_{T_{\alpha}}W=K_{T_{\alpha}}W = -2\frac{v^\alpha}{\KK}\ W\ , \nonumber \\
  D_{G^a}W&=& K_{G^a}W=2i G_{ab}b^b\ W\ , \qquad
  D_{z^{k}}W=i\int \chi_{k} \wedge G_3 \ ,
  \label{kcov1}
\end{eqnarray}
where we  used that $\chi_{ k}$ has the additional property \cite{CdO}
\begin{equation}
  D_{z^{ k}} \Omega =  i\chi_{ k}\ .
  \label{def-chi}
\end{equation}
Due to the fact that $K_k$ satisfies the no-scale condition \eqref{NScond}
the potential is unchanged if one includes $b^a$ and $c^a$. 
Inserting \eqref{kcov1} and \eqref{NScond} into 
\eqref{N=1pot} one shows that \eqref{potential37}  is reproduced.

It is interesting to note that both $W(\tau,z^k)$ and $f_{\kappa\lambda}(z^k)$
depend on the complex structure deformations $z^k$ but not
on the K\"ahler deformations. This can be directly understood from 
the Peccei-Quinn symmetries of the theories or equivalently
from the fact that the K\"ahler deformations can be viewed
as members of linear multiplets. Let us turn to this aspect now.

%
%
\subsection{The effective action in terms of linear multiplets}\label{Linearm}

In section~\ref{o3-spectrum} (table~\ref{N=1spectrum})
we already observed that for the 
$h^{(1,1)}_+$ chiral multiplet $(v^\alpha,\rho_\alpha)$ including the
K\"ahler deformations $v^\alpha$ there exist an dual description in terms 
of linear multiplets. 
In this section we rewrite the effective action using the linear multiplet
formalism of ref.\ \cite{BGG}. 
In this way we will be able to understand the definition of the K\"ahler
coordinates given in \eqref{tau} as a superfield duality transformation
and furthermore discover the no-scale property \eqref{NScond} of $K_k$ 
somewhat more conceptually. In an analog three-dimensional situation this has 
also been observed in \cite{BHS}.

Let us first briefly review $N=1$ supergravity coupled to
 $h^{(1,1)}_+$ linear multiplets 
$L^\alpha, \alpha=1,\ldots, h^{(1,1)}_+$ and
$r$ chiral multiplets $N^A, A=1,\ldots,r$ following \cite{BGG}.
Linear multiplets are defined by the constraint 
\beq\label{linearc}
(D^2-8\bar R) L^\alpha = 0 = (\bar D^2-8R) L^\alpha\ ,
\eeq
where $D$ is the superspace covariant derivative and $R$ is the chiral 
superfield containing the curvature scalar.
As bosonic components $L$ contains a  real scalar field which we also
denote by $L$ and the field strength of a
two-form $D_2$.
The superspace Lagrangian (omitting the gauge interactions) is given by 
\beq\label{actionL}
\cL = - 3 \int   E\, F(N^A,\bar N^A, L^\alpha) 
+ \frac12 \int \frac{E}{R}\, e^{K/2}\ W(N)
+ \frac12 \int \frac{E}{R^\dagger}\, e^{K/2}\ \bar W(\bar N)
\ ,
\eeq
where $E$ is the super-vielbein and $W$ the superpotential.
The function $F$
 depends implicitly  on the K\"ahler potential 
$K(N^A,\bar N^A, L^\alpha)$
through the differential constraint\footnote{Strictly speaking
$K(N^A,\bar N^A, L^\alpha)$ is not a K\"ahler potential 
but as we will see it determines the kinetic terms in the action.}
\bea\label{Fcon}
 1- \frac{1}{3}L^\alpha K_{L^\alpha}  = F-L^\alpha F_{L^\alpha}\ ,
\eea
which ensures the correct normalization of the Einstein-Hilbert term.
The subscripts on $K$ and $F$ denotes differentiation, i.e.\
$K_{L^\alpha} = \frac{\partial K}{\partial L^\alpha},
F_{L^\alpha} = \frac{\partial F}{\partial L^\alpha}$, etc.\ .
 
Here we are not interested in the most general couplings
but our aim is to rewrite the action \eqref{S_scalar}
in the linear multiplet formalism. As we are going to show 
this is achieved by the K\"ahler potential
\beq\label{KL}
  K = K_0(N^A, \bar N^{A}) + \alpha\ln (\KK_{\alpha \beta \gamma} L^\alpha L^\beta L^\gamma)\ ,
\eeq
where we
leave $K_0(N^A, N^{\bar A})$ and the normalization constant
$\alpha$ arbitrary for the moment.
Inserting \eqref{KL} into \eqref{Fcon} determines $F$ to be
\beq\label{FL}
   F=1 - \alpha + L^\alpha \zeta^R_\alpha(N^A, \bar N^{ A}) \ ,
\eeq
where  the real functions $\zeta^R_\alpha(N^A, \bar N^{A})$ are
not further determined by \eqref{Fcon}. In that sense the 
$\zeta^R_\alpha(N^A, \bar N^{A})$ are additional input functions
which determine the Lagrangian.

The (bosonic) component Lagrangian derived from \eqref{actionL} 
is found to be\footnote{This is a straightforward generalization
of the Lagrangian for one linear multiplet given in \cite{BGG}.
The potential for this case has also been given in \cite{HL}.}
\bea\label{kinetic}
\cL &=& -\frac{1}{2}R*\mathbf{1} - 
  \tilde K_{A\bar B}\, dN^A \wedge * d \bar N^{B}
  + \frac{1}{4} K_{L^\alpha L^\beta}\, 
  dL^\alpha \wedge * dL^\beta - V \nn\\ 
  && + \frac{1}{4} K_{L^\alpha L^\beta}\, dD^\alpha_2 \wedge * dD^\beta_2
     +  \frac{3i}2\,  dD^\alpha_2 \wedge 
\big(\zeta^R_{\alpha,A}\,dN^A -\zeta^R_{\alpha,\bar A}\,d\bar N^A\big)
\ ,
\eea
where 
\beq\label{Lsc}
\tilde K_{A\bar B} \equiv K_{A\bar B}- 3 L^\alpha \zeta^R_{\alpha,A\bar B}
\ , \qquad
 V = e^K \Big(\tilde K^{A \bar B}D_AW D_{\bar B}\bar W - 
(3- L^\alpha K_{L^\alpha}) |W|^2  \Big)\ .
\eeq
We see that the effective action is determined in terms of 
$K(N,\bar N, L)$ and $\zeta_\alpha^R(N,\bar N)$. 
$K$ determines the kinetic terms of the fields $N^A$ and $L^\alpha$ while  
the $\zeta_\alpha^R(N,\bar N)$ determine 
the couplings of the two-forms $D^\alpha_2$ to the chiral fields $N^I$.
Note that only derivatives of  $\zeta_\alpha^R$ appear leaving a 
constant piece in  $\zeta_\alpha^R$ undetermined.

From here we can proceed in two ways.
We can dualize the two-forms $D^\alpha_2$ 
in components and show the equivalence with the action \eqref{S_scalar}.
This is done at the end of this section.
However, performing the duality in superspace yields 
directly the proper K\"ahler coordinates $T_\alpha$ and 
thus gives a more conceptual understanding of the 
definition \eqref{tau}.
 
The duality transformation in superfields is 
performed in detail in \cite{BGG} and here we only repeat the 
essential steps.
One first considers  the linear multiplets $L^\alpha$ to be 
unconstrained real superfields and modifies the action
\eqref{actionL} to read\footnote{We omit the superpotential
terms here since they only depend on $N$ and play no role
in the dualization.}
\beq\label{actionX}
S = - 3 \int E\, \Big(F(N^A,\bar N^A, L^\alpha) + 
    \tfrac{2}{3} L^\alpha(T_\alpha + \bar T_\alpha) \Big) + \ldots\ ,
\eeq
where the $T_\alpha$ are chiral superfields and in order to be consistent
with our previous conventions we have included a factor $\frac{2}{3}$
in the second term. 
Variation with respect to $T_\alpha$ results in the constraint that $L^\alpha$ are linear multiplets
and one arrives back at the action \eqref{actionL}. 
Variation with respect to the (unconstrained) $L^\alpha$ yields the 
equations\footnote{Notice that there is a misprint
in the equivalent equation given in \cite{BGG}. We thank R.\ Grimm for
discussions on this point.}
\beq \label{bGl}
  \tfrac{2}{3} (T_\alpha + \bar T_\alpha)  + F_{L^\alpha}
- \tfrac{1}{3} K_{L^\alpha} 
\big(F+ \tfrac{2}{3} L^\beta (T_\beta + \bar T_\beta)\big)  =0 \ ,
\eeq
where we have used 
$\delta_{L} E = -\tfrac{1}{3} E K_{L^\alpha} \delta L^\alpha$.
This equation determines  
$L^\alpha$ in terms of the chiral superfields $N^A,T_\alpha$ and is the looked
for duality relation.
However, depending on the specific form of $F$ and $K$ 
one might not be able to solve \eqref{bGl} explicitly
for $L^\alpha$ but instead only obtain an implicit
relation  $L^\alpha(N,\bar N, T+\bar T)$.
Nevertheless one should 
insert  $L^\alpha(N,\bar N, T+\bar T)$ back  into \eqref{actionX} 
which then expresses the Lagrangian (implicitly) in terms 
of $T_\alpha$ and therefore defines a Lagrangian in the chiral superfield
formalism. 
The unusual feature being that the explicit functional dependence is 
not known.

A correctly normalized Einstein-Hilbert term is ensured by 
 additionally imposing
\bea \label{normeq}
  F(N,\bar N,L) + \tfrac{2}{3} L^{\alpha}(T_\alpha + \bar T_\alpha) = 1\ .
\eea 
Contracting \eqref{bGl} with $L^\alpha$ and using equation \eqref{normeq} one obtains
\eqref{Fcon}. Thus $F$
has to have the same functional dependence as before
and therefore eqs.\ \eqref{KL} and \eqref{FL} are unmodified but 
one should insert $L(N,\bar N,T+\bar T)$ implicitly determined by \eqref{bGl}.
In other words eqs.\ \eqref{KL} should be red as
\beq\label{KX}
 K = K_0(N,\bar N) +   \alpha \ln \Big[\KK_{\alpha \beta \gamma} \,
L^\alpha(N,T)\,   L^\beta(N,T)\,  L^\gamma(N,T)\Big]\ .
\eeq
Inserting \eqref{normeq} into \eqref{bGl} 
and using \eqref{KX} we arrive at
\beq \label{restr}
   T_\alpha + \bar T_\alpha +\tfrac{3}{2} \zeta^R_\alpha = \tfrac{1}{2} K_{L^\alpha}\ . 
\eeq
Comparing \eqref{KX} and \eqref{restr}
with \eqref{tau} and \eqref{zetadef} we are led to identify\footnote{Strictly
speaking \eqref{restr} only determines the real part
of $\zeta_\alpha$. However, the imaginary part can be redefined into 
the imaginary part of $T_\alpha$. We return to this point at the end of this 
section.}
\beq\label{zetaid}
\alpha = 1\ , \qquad  L^\alpha = \frac{v^\alpha}{\KK} \ , \qquad 
 \zeta^R_\alpha = \zeta_\alpha + \bar\zeta_\alpha\ , \qquad 
\zeta_\alpha =-\frac{i}{2(\tau-\bar \tau)}\ \KK_{\alpha b c}G^b (G- \bar G)^c\ .
\eeq
As promised we just showed that 
the somewhat ad hoc definition of the K\"ahler coordinates in \eqref{tau}
is nothing but the duality relation \eqref{restr} obtained from the superfield
dualization of  the linear multiplets $L^\alpha$ 
to chiral multiplets $T^\alpha$.

The case $\alpha=1$ is a somewhat special situation 
in that the function $F$ does not have a constant piece but only the term
linear in $L^\alpha$.
This in turn requires that the $\zeta_\alpha$ cannot be chosen zero but that they
have at least a constant piece so that $F$ does not
vanish. This constant is otherwise irrelevant since it
drops out of all physical quantities.\footnote{We 
thank R.\ Grimm for discussions about this point.}
(In a slightly different context 
the case $\alpha=1$ has also been discussed  in ref.\ \cite{Binetruy}.)

The discussion so far is valid for arbitrary $K_0$. By comparing
the kinetic terms of the chiral fields
in \eqref{kinetic} one determines 
\beq\label{K0}
K_0 =  K_{\rm cs}(z,\bar z)  -\text{ln}\big[-i(\tau - \bar \tau)\big] \ .
\eeq
Inserting \eqref{zetaid} and \eqref{K0}  into \eqref{KX} we finally arrive at
\beq \label{kaehlerpot-O7-2}
  K =  K_{\rm cs}(z,\bar z)  -\text{ln}\big[-i(\tau - \bar \tau)\big]   
- 2 \ln \big[\KK_{\alpha\beta\gamma}v^\alpha(T,\tau,G)\, v^\beta(T,\tau,G)\, v^\gamma(T,\tau,G)\big] \ ,
\eeq
where $v^\alpha(T,\tau,G)$ is determined by \eqref{restr}. This $K$ indeed agrees with 
the $K$ previously determined in \eqref{kaehlerpot-O7-1}.

Let us summarize the story so far.
$N=1$ supergravity coupled to 
chiral and linear multiplets is determined (in the formalism of ref.\
\cite{BGG} and apart from $W$ and $f$ which we can neglect for this discussion)
by the independent functions $K$ and $\zeta_\alpha$. 
For the orientifold compactification under consideration 
the chiral multiplets are $(z^k,\tau,G^a)$, the linear multiplets are
$L^\alpha$ and we determined
\bea\label{Ksum}
 K(z,\tau,L) &=& K_{\rm cs}(z,\bar z)  -\text{ln}\big[-i(\tau - \bar \tau)\big]
+   \ln \big[\KK_{\alpha \beta \gamma} \,
L^\alpha\,   L^\beta\,  L^\gamma\big]\ , \\
  F(L,\tau,G) &=& L^\alpha \, (\zeta_\alpha+\bar \zeta_\alpha)\ , \qquad
\zeta_\alpha =-\frac{i}{2(\tau-\bar \tau)}\ \KK_{\alpha b c}G^b (G- \bar G)^c
+{\rm const.}  \ .      \nn
\eea
In the dual formulation where the linear multiplets $L^\alpha$ are dualized
to chiral multiplets $T_\alpha$ the Lagrangian is entirely determined
by the K\"ahler potential given in \eqref{kaehlerpot-O7-2}  with the `unusual'
feature that it is not given explicitly in terms of the chiral
multiplets but only implicitly via the constraint \eqref{restr}.
In this sense the orientifold compactifications 
(and similarly the compactifications of F-theory on elliptic Calabi-Yau
fourfolds considered in \cite{HL}) lead to 
a more general class of K\"ahler potentials
then usually considered in supergravity.
In fact the same feature holds for arbitrary $K_0$ and arbitrary $\zeta_\alpha$.

Furthermore, these `generalized' K\"ahler potentials are all of 
`no-scale type' in that they lead to a positive semi-definite potential $V$.
For $\alpha=1$ (and arbitrary $K_0$ and $\zeta_\alpha$) 
the K\"ahler potential \eqref{KL} obeys 
\beq
L^\alpha K_{L^\alpha} = 3\ ,
\eeq
and hence the
the second term in the potential \eqref{Lsc} vanishes leaving a positive semi-definite
potential with a supersymmetric Minkowskian ground state.
Since in the chiral formulation $K$ cannot even be given explicitly one can
consider such $K$s  as a `generalized' class of 
no-scale K\"ahler potentials.\footnote{It would be interesting 
to show that all positive 
semi-definite potentials can be derived from such
K\"ahler potentials involving linear multiplets and in this way offering a
classification of non-scale K\"ahler potentials.} 
The analogous property has also been observed in refs.\ \cite{HL,BBHL,DAFT}.
Finally note with what ease the no-scale property follows in the 
linear formulation compared to the somewhat involved computation
in the chiral formulation performed in appendix~\ref{Kmetrics3}.

 To complete our analysis we also perform the duality transformation 
 in components, i.e.\  we dualize the two-forms $D^\alpha_2$ to scalars
 $\rho_\alpha$. This can be done by the standard procedure \cite{BGG}
where one replaces  $dD_2^\alpha$ in \eqref{kinetic}
by an unconstrained three-form $D_3^\alpha$ and in addition adds the term
 \beq
   \delta \mathcal{L} = -3\, D_3^\alpha \wedge d\rho_\alpha^\prime\ , 
 \eeq
 to the Lagrangian \eqref{kinetic}. The $\rho_\alpha^\prime $ act
 as Lagrange multipliers in that their equations of motions 
imply $D^\alpha_3=dD_2^\alpha$. On the other hand the equations
of motion for $D_3^\alpha$ are now algebraic and can be used to eliminate
 $D_3^\alpha$ in favor of $\rho_\alpha^\prime$. They are given by
  \beq\label{D3}
   D^\alpha_3 = 6\, K^{L^\alpha L^\beta} 
   * \big(d\rho_\beta - i (\bar \zeta_{\beta,A} dN^A -  \zeta_{\beta,\bar A} d\bar N^A) \big)\ ,
 \eeq
where in order to make contact with \eqref{scalaract} we have 
further defined 
$\rho_\alpha^\prime=\rho_\alpha +\frac{i}{2} (\zeta_\alpha-\bar \zeta_\alpha )$.
Inserting \eqref{D3} into $\cL+\delta \cL$ one finally obtains
 \bea \label{dualtoL}
  \cL &=& -\frac{1}{2}R*\mathbf{1} - 
   (K_{A\bar B} - 3 L^\alpha \zeta^R_{\alpha,A\bar B})\, dN^A \wedge * d \bar N^{B}
   + \frac{1}{4} K_{L^\alpha L^\beta}\, 
   dL^\alpha \wedge * dL^\beta \\ 
   && + 9 K^{L^\alpha L^\beta} 
   \big(d\rho_\alpha - i (\bar \zeta_{\alpha,A} dN^A -  \zeta_{\alpha,\bar A} d\bar N^A) \big)\wedge 
   * \big(d\rho_\beta - i (\bar \zeta_{\beta,A} dN^A -  \zeta_{\beta,\bar A} d\bar N^A) \big)\ . \nn
 \eea
 For  $\alpha=1$  and $L^\alpha = \frac{v^\alpha}{\KK}$ 
the metric in  \eqref{dualtoL} coincides with the metric
 \eqref{scalaract} obtained in the chiral description. 

Let us close this section by stressing once more that the analysis just
performed
can easily be generalized to other situations in particular including
background $D$-branes. For example, $D3$-branes  couple to $\hat C_4$ 
and therefore also to $D^\alpha_2$. 
As shown in \cite{GGJL} the low energy dynamics and 
couplings of the $D3$-brane scalars is 
fully encoded in $\zeta_\alpha$ and $K(z,\tau,L)$ remains unchanged. 
However, in the dual formulation with chiral multiplets
the K\"ahler potential \eqref{KX} changes due a modified \eqref{restr}. 
By differentiating \eqref{restr} with respect 
to the chiral fields $T_\alpha, N^A$ one obtains
\beq
  \frac{\partial L^\alpha(T,N)}{\partial T_\beta} = 2 K^{L^\alpha L^\beta}\ , \qquad 
  \frac{\partial L^\alpha(T,N)}{\partial N^A} = 3 K^{L^\alpha L^\beta} \zeta^R_{\alpha, A}\ , 
\eeq 
which determines the K\"ahler metric
even so  $L^\alpha(T,N)$ is not given explicitly.


\section{Calabi-Yau orientifolds with $O5/O9$ planes \label{O5O9-planes}}

In this section we repeat the analysis of section~\ref{O3O7-planes}
for orientifold compactifications which satisfy
the second possible projection \eqref{o5-projection} which reads 
$\OO_{(2)}=\Omega_p\sigma^*$ with $\sigma^*\Omega=\Omega$.
As we already discussed earlier, in this case the possible orientifold
planes present are $O5$- or $O9$-planes. 
As in this previous section let us start by imposing
\eqref{o5-projection} on the ten-dimensional massless spectrum and 
in this way determine the massless $N=1$ supermultiplets in $D=4$.

\subsection{The massless spectrum}

In order to determine the invariant spectrum we first recall the 
transformation of 
the ten-dimensional massless fields under the world-sheet parity 
transformation
$\Omega_p$. As we discussed in section~\ref{O3O7-planes}
the dilaton $\hat \phi$, the metric $\hat g$ and 
the two form $\hat C_2$
are even while 
$\hat l$,
$\hat B_{2}$ and  $\hat C_4$ are odd.
Using \eqref{o5-projection} this implies that the invariant
states have to obey
\begin{equation}
\begin{array}{lcl}
\sigma^*  \hat \phi &=& \  \hat \phi\ , \\
\sigma^*   \hat g &=& \ \hat g\ , \\
\sigma^*   \hat B_2 &=& -  \hat B_2\ ,
\end{array}
\hspace{2cm}
\begin{array}{lcl}
\sigma^*   \hat l &=&  -  \hat l\ , \\
\sigma^*   \hat C_2 &=& \  \hat C_2\ ,  \\
\sigma^*  \hat C_4 &=&  -  \hat C_4\ .
\end{array}
\label{fieldtransf2}
\end{equation}
In addition, $\sigma^* $ now is required to satisfy
\beq\label{Omegatrans2}
\sigma^*\Omega =\Omega\ .
\eeq 

As before $\sigma$ has to be a holomorphic isometry of $Y$.
Therefore the K\"ahler form is invariant and eq.\ \eqref{transJo}
remains unchanged implying that $h_+^{(1,1)}$ K\"ahler deformations
$v^{\alpha}(x)$ survive the projection.
However, due to \eqref{Omegatrans2} instead of \eqref{Omegatrans}  
the deformations
of the complex structure change in that now $h_+^{(1,2)}$
complex structure deformation $z^\kappa$ survive the projection
or in other words 
\eqref{cso} is replaced by 
\beq\label{csoO5}
\delta{g}_{ij} =  \frac{i}{||\Omega||^2}\, \bar z^{\kappa} 
(\bar \chi_{ \kappa})_{i\ib\bj}\,
\Omega^{\ib\bj}{}_j \ , \quad \kappa=1,\ldots,h_+^{(1,2)}\ ,
\eeq
where $\bar\chi_{\kappa}$ denotes a basis of $H^{(1,2)}_+$.

From eqs.\ \eqref{fieldtransf2} we learn that in the expansion 
of $\hat B_2$ the odd elements survive while for 
$\hat C_2$ and $\hat C_4$ the even elements are kept.
Thus the expansion \eqref{exp1} is replaced by
\bea\label{expO5} 
\hat B_2 &=& b^a(x)\ \omega_a\ , \qquad a=1,\ldots, h_-^{(1,1)}\ ,\nn\\
\hat C_2& =& C_2(x)+ c^\alpha(x)\ \omega_\alpha\ , 
\qquad \  \alpha = 1,\ldots,h_+^{(1,1)}\ , \\
\hat{C}_4 &=& D_{2}^a(x) \wedge \omega_a + V^{k}(x) \wedge 
\alpha_{k} - U_{k}(x) \wedge \beta^{k} + 
  \rho_a(x)\, \tilde \omega^a\ , \qquad k=1,\ldots,h_-^{(1,2)}\ .  
\nonumber 
\eea
As before the self-duality constraint 
$*\hat F_5=\hat F_5$ eliminates half of the degrees of freedom 
in $\hat C_4$. Again we have the choice to eliminate $ D_{2}^a(x)$
or $\rho_a$ and express the action in terms of chiral or linear multiplets.
Similarly,  the 
axion $\hat l$ is now projected out and replaced by the $D=4$ antisymmetric
tensor $C_2(x)$. As a consequence  the $N=1$ spectrum contains a
`universal' linear multiplet which in the massless case can be dualized 
to a chiral multiplet.

Altogether the resulting $N=1$ spectrum assembles into
a gravitational
multiplet, $h_-^{(2,1)}$ vector multiplets and 
either $(h_+^{(2,1)}+ h^{(1,1)}+1)$ chiral multiplets 
or $(h_+^{(2,1)}+ h^{(1,1)}_+)$ chiral multiplets 
and $(h^{(1,1)}_-+1)$ linear multiplets. We summarize
the spectrum in  table~\ref{N=1spectrum2} \cite{BH}.

\begin{table}[h] \label{N=1spectrum2}
\begin{center}
\begin{tabular}{|l|c|c|} \hline 
 \rule[-0.3cm]{0cm}{0.8cm} 
 gravity multiplet&1&$g_{\mu \nu} $ \\ \hline
 \rule[-0.3cm]{0cm}{0.8cm} 
 vector multiplets&   $h_-^{(2,1)}$&  $V^{k} $\\ \hline
 \rule[-0.3cm]{0cm}{0.8cm} 
 \multirow{2}{30mm}[-2.0mm]{chiral multiplets} &   $h_+^{(2,1)}$& 
$z^{\kappa} $ \\ \cline{2-3}
 \rule[-0.3cm]{0cm}{0.8cm} 
 &     $h^{(1,1)}_+$& $( v^\alpha, c^\alpha )$ \\ \hline
 \rule[-0.3cm]{0cm}{0.8cm} 
\multirow{2}{30mm}[-2.5mm]{chiral/linear multiplets} 
 &  $ h^{(1,1)}_-$ &$( b^a, \rho_a)$ \\ \cline{2-3}
 \rule[-0.3cm]{0cm}{0.8cm} 
   & 1 & $(\phi,C_2)$ \\ \hline
\end{tabular}
\caption{$N =1$ spectrum of $O5/O9$-orientifold compactification.}
\end{center}
\end{table} 

Compared to the spectrum of the first projection given in 
table~\ref{N=1spectrum} we see that the vectors and complex structure 
deformations have switched their role with respect to the decomposition
in $H^{(3)}$. Furthermore, different real fields combine into the complex
scalars of the chiral/linear multiplets or in other words the complex
structure on the moduli space has changed. Now 
$(v, c)$ and $( b, \rho)$ combine into chiral multiplets
whereas before $(v, \rho)$ and $(b, c)$ formed the 
chiral multiplets.\footnote{Note that again the complex structure which
combines $(v, b)$ and which is natural from the $N=2$ Calabi-Yau
point of view does not appear.}

\subsection{The effective action \label{O5effa}}

To evaluate the four-dimensional effective action we proceed as in the $O3/O7$ case, by
first evaluating the field strengths (\ref{fieldstr}) including 
the possibility of background three-form  fluxes 
$H_3$ and $F_3$.
Since $\hat B_2$ and hence  $H_3$ is odd
it is again parameterized by $H^{(3)}_-$ while $\hat C_2$ and 
$F_3$ are even and therefore parameterized by $H^{(3)}_+$.
Due to \eqref{Omegatrans2} there is a slight change in $H^{(3)}$
compared to the previous case in that 
now we have $h^{(3,0)}_+=h^{(0,3)}_+= 1$ and $h^{(3,0)}_-=h^{(0,3)}_-=0$
or in other words we have  the split
$H^{(3)}_+ = H^{(0,3)}_+\oplus H^{(1,2)}_+ 
\oplus H^{(2,1)}_+ \oplus H^{(3,0)}_+$ and 
$H^{(3)}_- = H^{(1,2)}_- \oplus H^{(2,1)}_-$.
As a consequence the 
explicit expansions of the background fluxes $H_3$ and $F_3$
are given by
\bea\label{mef}
H_3&=& m^k_H \alpha_k - e_k^H \beta^k\ , \qquad k = 1, \ldots, h^{(2,1)}_-\ ,
\nn\\
F_3 &=& m_F^{\hat \kappa} \, \alpha_{\hat \kappa} 
- e^F_{\hat \kappa}\, \beta^{\hat \kappa} \ ,
\qquad \hat\kappa = 0, \ldots, h^{(2,1)}_+ \ ,
\eea
where the $(m^k_H, e_k^H)$ are $2h^{(2,1)}_-$ constant flux 
parameters determining $H_3$ and 
$(m_F^{\hat \kappa}, e^F_{\hat \kappa})$ are $2h^{(2,1)}_++2$
constant flux 
parameters corresponding to $F_3$. 
Inserting \eqref{expO5} and  \eqref{mef} into \eqref{fieldstr}
we obtain
\bea
 \hat H_3 &=& db^a \wedge \omega_a + m^k_H \alpha_{k} - 
             e_{k}^H \beta^{k}\ ,\qquad
 \hat F_3 \ = \ dC_2+dc^\alpha \wedge \omega_\alpha + 
m_F^{\hat \kappa} \, \alpha_{\hat \kappa} 
- e^F_{\hat \kappa}\, \beta^{\hat \kappa}
,\nonumber \\
 \hat F_5 &=& dD_2^a \wedge \omega_a + \tilde F^{k} \wedge \alpha_{k}
               - \tilde G_{k}\wedge \beta^{k} 
               + d\rho_a \wedge \tilde \omega^a \nonumber\\
           && - db^a \wedge C_2 \wedge \omega_a - c^\alpha db^a \omega_a \wedge 
               \omega_\alpha\ ,
\eea
where we defined
\beq\label{Fcech}
\tilde F^{k}= dV^{k} - m^{k}_H C_2\ , \qquad
\tilde G_{k}=dU_{k} - e_{k}^H C_2\ .
\eeq 
As before the self-duality condition on $\hat F_5$
is imposed by adding to the action the total derivative \cite{DallAgata}
\begin{equation}
  \delta S^{(4)}_{O5/O9} = \frac{1}{4} dV^{k} \wedge dU_{k} + 
  \frac{1}{4} dD_2^a \wedge d\rho_a\ .
  \label{totalderiv2} 
\end{equation} 
Again we have the choice which fields to eliminate. In order to make contact
with the standard $N=1$ supergravity we first eliminate 
$D^a_{2}$ and $U_{k}$ by inserting their equations
of motion into the action.
After Weyl rescaling the four-dimensional metric with a factor $\KK/6$ 
the ${N}=1$ effective action reads 
\bea\label{actiono5}
 S^{(4)}_{O5/O9} &=& \int -\frac{1}{2}R*\mathbf{1}-
  G_{\kappa\lambda} \; dz^{\kappa} \wedge *d\bar z^{\lambda}
  -G_{\alpha \beta} \; dv^\alpha \wedge *dv^\beta 
   - \frac{1}{4}d\, \text{ln} \KK \wedge * d\, \text{ln} \KK \nonumber \\ 
  &&-\frac{1}{4} d\phi \wedge * d\phi
  - e^{\phi} G_{\alpha \beta}\; dc^\alpha \wedge * dc^\beta
  - e^{-\phi} G_{ab}\; db^a \wedge * db^b \nonumber \\ 
  && - \frac{\KK^2}{144}\, e^{\phi} dC_2 \wedge * dC_2
  -\frac{1}{4} dC_2 \wedge (\rho_a db^a - b^a d\rho_a)\nonumber \\
  &&- \frac{9}{4\KK^2}\, G^{ab}(d \rho_a - \KK_{ac\alpha} c^\alpha db^c)
  \wedge *(d \rho_b - \KK_{bd\beta} c^\beta db^d)- V*\mathbf{1}\\
&& 
 + \frac{1}{4} \text{Re}\; \mathcal{M}_{k l}\; \tilde 
  F^{k} \wedge \tilde F^{l} + 
  \frac{1}{4} \text{Im}\; \mathcal{M}_{k l}\; 
  \tilde 
  F^{k} \wedge * \tilde F^{l} +
  \frac{1}{4} e_{k} (dV^{k} +
  \tilde F^{k})\wedge C_2\nonumber
\eea
where 
\bea\label{pot5}
V &=&
  \frac{18i\ e^{\phi}}{\KK^2 \int \Omega \wedge \bar \Omega }
  \left( \int \Omega \wedge F_3 \int \bar \Omega \wedge F_3 
  + G^{\kappa \lambda} \int \chi_{\kappa} \wedge F_3 
  \int \bar \chi_{\lambda} \wedge F_3 \right) \\
  &-& 
   \frac{9}{\KK^2}e^{-\phi} \Big[
  m^{k}_H\, (\text{Im} \MM)_{ k l}\,   m^{l}_H +
  \big(e_{k}^H-(m_H \text{Re} \MM)_{k}\big)   \big(\text{Im} \MM\big)^{-1  k l} 
  \big(e_{l}^H-(m_H \text{Re} \MM)_{l}\big) \Big]\ . \nonumber
\eea
The derivation of this potential is performed in 
appendix~\ref{Spotentials5} where we also show that the first term
in the potential can be written in terms of 
$(e^F_{\hat\kappa},m_F^{\hat\kappa})$ 
exactly as the second term.\footnote{Note that in this class of 
orientifolds the topological term 
$\int_YH_3 \wedge F_3$ vanishes since there is no intersection between 
$H^{(3)}_+$ and $H^{(3)}_-$. Thus strictly speaking background
D-branes have to be included in order to satisfy the tadpole cancellation
condition.} 

The action \eqref{actiono5} has the standard one-form gauge invariance
$V^k\to V^k+d\Lambda^k_0$ 
but due to the modification in \eqref{Fcech} 
also a modified (St\"uckelberg) two-form gauge invariance given by
\beq\label{2gauge}
C_2\to C_2 +d\Lambda_1\ ,\qquad V^k\to V^k + m^k_H\Lambda_1\ .
\eeq
Thus for $m^k_H\neq 0$ one vector can be set to zero by an appropriate 
gauge transformation. 
This is directly related to the fact that  \eqref{actiono5}
includes mass terms  proportional to $m^k_H$ for $C_2$ arising from 
\eqref{Fcech}. In this case gauge invariance requires
the presence of Goldstone degrees of freedom which
can be `eaten' by $C_2$.\footnote{Exactly the same situation
occurs in Calabi-Yau compactifications of type IIB
with background fluxes where both $B_2$ and $C_2$
can become massive \cite{LM}.}
Finally note that the last term in \eqref{actiono5} also 
includes a standard $D=4$ Green-Schwarz term $F^k\wedge C_2$.

\subsubsection{Vanishing magnetic fluxes $m^k_H=0$}

The next step is to show that $S^{(4)}_{O5/O9}$ is consistent 
with the constraints of $N=1$ supergravity. However, due to the
possibility of $C_2$ mass terms this is not completely
straightforward. A massive $C_2$
is no longer dual to a scalar but rather to a vector.
We find it more convenient to keep the 
massive tensor in the spectrum and discuss 
the $N=1$ constraints in terms of a massive linear multiplet.
However, in this case a `standard' $N=1$ action is not
available in the literature and thus we first discuss
the situation $m^k_H= 0$  where $\tilde F^{l}= F^{l}$ holds.
In this case the $C_2$ remains massless 
and  can be dualized to a scalar field $h$ which together with
the dilaton $ \phi$ combines to form a chiral multiplet 
$( \phi, h)$.
Using the standard dualization procedure (see, for example,  \cite{LM})
one obtains 
\begin{eqnarray}  \label{4d-action2} 
  S^{(4)}_{O5/O9}&=&\int -\frac{1}{2}R*\mathbf{1}-
  G_{\kappa \lambda} \; dz^{\kappa} \wedge *d\bar z^{\lambda}
  -G_{\alpha \beta} \; dv^\alpha \wedge *dv^\beta \nonumber \\ 
  && - \frac{1}{4}d\, \text{ln} \KK \wedge * d\, \text{ln} \KK
  -\frac{1}{4} d \phi \wedge * d\phi \nonumber \\
  && -e^{\phi} G_{\alpha \beta}\; dc^\alpha \wedge * dc^\beta
  - e^{-\phi} G_{ab}\; db^a \wedge * db^b \\ 
  &&- \frac{9}{\KK^2}e^{-\phi}(Dh+\frac{1}{2}(d\rho_a b^a -\rho_a db^a))
  \wedge *(Dh+\frac{1}{2}(d\rho_a b^a - \rho_a db^a))\nonumber \\
  &&- \frac{9}{4\KK^2}G^{ab}(d \rho_a - \KK_{ac\alpha} c^\alpha db^c)
  \wedge *(d \rho_b - \KK_{bd\beta} c^\beta db^d)\nonumber \\
  &&+\frac{1}{4}\text{Im}\; \cM_{kl}\; 
     dV^{k}\wedge *dV^{l}
     +\frac{1}{4}\text{Re}\; \cM_{kl}\;
     dV^{k}\wedge dV^{l}- V*\mathbf{1}\ , \nn
\end{eqnarray} 
where  $V$ is given by \eqref{pot5} evaluated at $m^k_H=0$
and the covariant derivative of $h$ is defined as
\beq\label{hcov}
Dh=d h - e_k^H V^k\ .
\eeq
$h$ couples non-trivially to the gauge fields as a direct consequence
of the Green-Schwarz coupling $F^k\wedge C_2$
in \eqref{actiono5}. In the dualized
action the scalar $h$ then is charged 
under the corresponding $U(1)$ gauge transformation. 
More precisely, the gauge invariance reads
\beq\label{ginv}
h\to h + e_k^H \Lambda_0^k\ ,\qquad
V^k \to V^k + d\Lambda_0^k\ ,
\eeq
which leaves the covariant derivative \eqref{hcov}
invariant. Note that the gauge charges are set by the electric fluxes.

In the action \eqref{4d-action2}  we immediately see that the complex
structure deformations $z^\kappa$ are again already good K\"ahler coordinates.
For the remaining fields we find the appropriate K\"ahler coordinates
to be 
\begin{eqnarray}\label{Kcoord}
  t^\alpha &=& \frac{\hat v^\alpha}{\hat \KK} + i c^\alpha\ , \nn \\
  A_a &=& -\frac{2}{3} \hat G_{ab} b^b 
    + i \left(\KK_{ab\alpha}c^\alpha b^b + \rho_a \right)\ ,\\
  S & = &   e^{\frac12\phi}\, \frac{\cK}6 + i h 
           - \frac{1}{4} (\text{Re}\N^{-1})^{ab} A_a (A+\bar A)_b\ , 
    \nonumber\\
    & = & e^{\frac12\phi}\,\frac{\cK}6  + i h + \frac{1}{3}\hat G_{ab} b^a b^b 
         - \frac{i}{2} \left( \KK_{ab \alpha} b^a b^b c^\alpha + \rho_a b^a \right)\ , \nn
\end{eqnarray}
where  we abbreviated
\beq\label{Ndef}
\N_{ab}(t) =-\frac{2}{3}\hat G_{ab} + i \KK_{ab \alpha}c^\alpha
=\KK_{ab \alpha} t^\alpha \ ,\qquad
  \hat v^\alpha = e^{\frac{1}{4} \phi} \KK^{-\frac{1}{2}} v^\alpha \ .
\end{equation}
Note that the quantities with the  hat (i.e.\ $ \hat \KK, \hat G_{ab}$)
in \eqref{Kcoord} are calculated using the redefined
K\"ahler moduli $\hat v^\alpha$.\footnote{In this case it is necessary
to redefine the K\"ahler moduli and the dilaton in analogy to Calabi-Yau 
compactifications of type IIB \cite{BGHL}.}
Furthermore, $\N_{ab}$ depends holomorphically on the coordinates $t^\alpha$
and the covariant derivative of $h$ given in \eqref{hcov}
translates into the covariant
$DS = dS - i  e_{k} V^{k}$.
In the variables given in \eqref{Kcoord} the K\"ahler potential reads
\begin{eqnarray}
  K &=& -\text{ln}\Big[-i\int\Omega \wedge \bar \Omega \Big]
        - \text{ln}\Big[\frac{1}{48}\KK_{\alpha \beta \gamma}(t+\bar t)^\alpha 
        (t+\bar t)^\beta (t+\bar t)^\gamma  \Big] \nonumber \\
    & & - \text{ln}\Big[S + \bar S + \frac{1}{4} (A + \bar A)_a (\text{Re}\N^{-1})^{ab} 
        (A + \bar A)_b \Big]\ . \label{O5-Kaehlerpot} 
\end{eqnarray}  
The corresponding K\"ahler metric is computed in appendix~\ref{Kmetrics5}
and the check
that it indeed reproduces \eqref{4d-action2} is straightforward, since 
\eqref{O5-Kaehlerpot} is closely related to the quaternionic
`K\"ahler potential' given in \cite{FS} and we can make use
of their results.\footnote{Note however,
that the complex structure changed non-trivially.
In \cite{FS} the standard $t \sim v + i b$ formed complex coordinates.}
 The same reference  already observed 
that for a holomorphic matrix $\N$ the quaternionic geometry is also 
K\"ahler. This situation was also found in compactifications
of the heterotic string to $D=3$ on a circle \cite{HL}.

Finally, let us note that 
inserting \eqref{Kcoord} back into $K$ results in 
\begin{equation}
  K = -\text{ln}\Big[-i\int\Omega \wedge \bar \Omega \Big]
      - \text{ln}\big[2 e^{-\phi}\big]  - 2\text{ln}\big[ \KK\big]\ ,
      \label{K-afterK}
\end{equation}
which is exactly the same $K$ as \eqref{kaehlerpot-O7-1} when expressed
in terms of the  variables $\phi$ and $v^\alpha$.
This can be understood from that fact that these NS-sector variables 
are the same in both types of orientifolds  and the difference 
in the two cases 
only arises when one expresses $K$ in terms of
proper K\"ahler coordinates.

Let us now turn to the gauge couplings and the potential. 
Comparing \eqref{actiono5}
with \eqref{N=1action} we determine
\begin{equation}\label{gaugekin-O5} 
  f_{kl }(z^\kappa)
=-\frac{i}{2}  \bar{\MM}_{kl} 
= -\frac{i}{2}\cF_{kl}\big|_{z^k=0=\bar z^k}\ ,
\end{equation}
where the second equation holds in complete analogy 
to eqs.\ \eqref{Fdiag}, \eqref{fholo}. As a consequence
the $f_{kl }$ are again manifestly holomorphic functions of
the complex structure deformations $z^\kappa$.

{}From eq.\ \eqref{ginv}
we see that the axion is charged 
and as a consequence we
expect a non-vanishing $D$-term in the potential. Recall
 the general formula
for the $D$-term \cite{WB}
\beq
K_{I\bar J} \bar X^{\bar J}_k = i \partial_I D_k \ ,
\eeq
where $X^{I}$ is the Killing vector of the $U(1)$ transformations
defined as $\delta M^I = \Lambda^k_0 X_k^J \partial_J M^I$.
Inserting \eqref{O5-Kaehlerpot}, \eqref{KmetricO5} 
and \eqref{Kcoord} we obtain 
\begin{equation}
  D_k = - e_{k}^H\, \frac{\partial K}{\partial \bar S} = 
  3\, e_{k}^H\, e^{-\frac12\phi}\cK^{-1} \ .
\end{equation}
Using also (\ref{gaugekin-O5}) we arrive at the $D$-term contribution 
to the potential 
\begin{eqnarray}
  \frac{1}{2}   (\text{Re}\; f)^{-1\ kl} D_k D_l = -
  9 e^{-\phi}\cK^{-2}\; e_{k}^H\, (\text{Im}\; \MM)^{-1\ kl}\,
  e_{l}^H\ , \label{D-term}
\end{eqnarray}
which indeed reproduces the last term in \eqref{pot5} for 
$m^k_H=0$.

The first term in \eqref{pot5} arises from the  superpotential
\begin{equation}\label{W5}
  W= \int_{Y} \Omega \wedge F_3\ .
\end{equation}
Using \eqref{O5-Kaehlerpot} the K\"ahler covariant derivatives 
of $W$ are calculated to be
\begin{eqnarray}\label{W5detail}
  D_{z^{\kappa}}W &=& i\int \chi_{\kappa} \wedge F_3 \ ,\qquad
  D_S W \ =\ K_S W = -3e^{-\frac12 \phi} \cK^{-1}  W  \ , \nonumber \\
  D_{t^\alpha}W &=&K_{t^\alpha} W = \frac{3}{2}\Big(K_\alpha 
  + e^{-\frac12 \phi} \cK^{-1}\KK_{\alpha a b}b^a b^b\Big) W \ , \\
  D_{A_a}W &=& K_{A_a}W= -3e^{-\frac12 \phi} \cK^{-1} b^a  W \ ,
\nonumber
\end{eqnarray}
where we used $D_{z^{\kappa}} \Omega = i\chi_{\kappa}$.
 Inserting \eqref{KinvO5} and \eqref{W5detail} into \eqref{N=1pot}
one obtains the first term in \eqref{pot5}.

It is interesting  that for this class of orientifolds
the RR-flux $F_3$ results in a contribution to the superpotential while
the NS-flux $H_3$ contributes instead to a $D$-term.

%
\subsubsection{Non-vanishing magnetic fluxes $m^k_H\neq 0$}

Let us now turn to the case where both electric and magnetic fluxes are
non-zero and the two-form $C_2$ is massive.
In this case $C_2$ is dual to a massive vector or equivalently the massive
linear multiplet is dual to massive vector multiplet.
Here we do not discuss this duality but instead show how the couplings
of a massive linear multiplet is consistent with the action 
\eqref{actiono5}.\footnote{More details will appear in \cite{schulgin}.}

In section~\ref{Linearm} we already reviewed some properties of the linear 
multiplet. The kinetic terms in the effective action are determined by
a generalized K\"ahler potential \eqref{KL} and the couplings $\zeta$ 
defined in \eqref{FL} which 
determine the couplings of the two-form to the chiral multiplets.
These
functions are not affected
by any  mass terms and can be red off directly from \eqref{actiono5}.
Comparing \eqref{actiono5} with \eqref{kinetic} and using 
\eqref{Kcoord} and \eqref{Ndef} we determine 
\bea\label{KpotL}
  K =  K_0 + \text{ln} L\ , \qquad
  \zeta^R =  \frac{1}{12} (A+ \bar A)_a (\text{Re}\N^{-1})^{ab} (A+ \bar A)_b\ 
\eea
with
\bea \label{KLdetail}
  K_0 &=& -\text{ln}\Big[-i\int\Omega \wedge \bar \Omega \Big] 
        - \text{ln}\Big[\frac{1}{48}\KK_{\alpha \beta \gamma}(t+\bar t)^\alpha 
        (t+\bar t)^\beta (t+\bar t)^\gamma  \Big]\ , \nonumber \\
  L &=& 3 e^{-\frac12\phi}\cK^{-1}\ , \qquad D_2 = \frac{1}{2} C_2  \ . \label{defLC}
\eea
Performing the duality transformation from the linear
multiplet $L$ to a chiral multiplet $S$ as outlined 
in section~\ref{Linearm} we obtain 
\bea
  \frac{1}{ L} = S + \bar S + 3\, \zeta^R\  .
\eea
Inserted back into \eqref{KpotL} we determine the dual K\"ahler potential 
to be
\bea
  K = K_0 - \text{ln}\Big[S + \bar S + 3\, \zeta^R (A, \bar A, t, \bar t)\Big]\ ,
  \label{dual-KP}  
\eea
which indeed coincides with \eqref{O5-Kaehlerpot}.
Thus we have shown that the kinetic terms can consistently
be described either in the chiral- or the linear multiplet formalism
and that \eqref{Kcoord} are the appropriate coordinates.

Let us now we briefly discuss the situation of a massive
linear multiplet coupled to $N=1$ vector- and chiral multiplets.
For simplicity we discuss the situation in flat space 
and do not couple the massive linear multiplet to supergravity.
However, we expect our results to generalize to the 
supergravity case. 
More details can be found in \cite{GGRS,schulgin}.

As we already said, a 
linear multiplet $L$ contains a real scalar (also denote by $L$)
and the field strength of
a two-form $C_2$ as bosonic components. However, 
it does not contain the two-form itself
which instead is a member of the chiral `prepotential' $\Phi$ defined
as\footnote{We suppress the spinorial indices and use the convention
$D\Phi \equiv D^\alpha\Phi_\alpha$, 
$\bar D\bar\Phi \equiv \bar D^{\dot{\alpha}}\bar\Phi_{\dot{\alpha}}$.}
\beq
L= D\Phi +\bar D \bar \Phi\ , \qquad \bar D\Phi = 0\ .
\eeq
This definition solves the constraint \eqref{linearc} (in flat space).
The kinetic term for $L$ (or rather for $\Phi$) is given in 
\eqref{actionL} and a mass-term can be added via the 
chiral integral
\beq\label{Lmasst}
\cL_{m} = \frac14\int d^2\theta \Big[
f_{kl}(N) (W^k - 2i m^k_H\Phi)(W^l - 2i m^l_H\Phi)
+ 2 e_k^H (W^k - i m^k_H\Phi)\Phi\Big] + {\rm h.c.}\ ,
\eeq
where $W^k= -\tfrac14 \bar D^2 DV^k$ are the chiral field strengths supermultiplets
of the vector multiplets $V^k$ and $f_{kl}(N)$ are the gauge kinetic function
which can depend holomorphically on the chiral multiplets $N$.
$(m^k_H,e_k^H)$ are constant parameters which will turn out 
to correspond to the flux parameters defined in \eqref{mef}.
The Lagrangian \eqref{Lmasst} is invariant under the standard
one-form gauge invariance $V^k\to V^k +\Lambda_0^k + \bar \Lambda_0^k$
($\Lambda_0^k$ are chiral superfields)
which leaves both $W^k$ and $\Phi$ invariant.
In addition \eqref{Lmasst} has a two-form gauge invariance 
corresponding to \eqref{2gauge} given by
\beq\label{linearg}
\Phi \to \Phi +\frac{i}8 \bar D^2 D \Lambda_1\ ,\qquad
V^k\to V^k + m^k_H \Lambda_1\ ,
\eeq
where $\Lambda_1 $ now is a real superfield. 
{}From \eqref{linearg} we see that one entire vector multiplet
can be gauged away and thus plays the role of the Goldstone degrees
of freedom which are `eaten' by the massive linear multiplet.

In components one finds the bosonic action
\beq
\cL_{m} = -\frac{1}{2} \text{Re} f_{k l}\; \tilde 
  F^{k} \wedge  * \tilde F^{l} -
  \frac{1}{2} \text{Im} f_{k l}\; 
  \tilde 
  F^{k} \wedge \tilde F^{l} +
  \frac{1}{4} e_{k} (dV^{k} +
  \tilde F^{k})\wedge C_2 - V*\mathbf{1} \ ,
\eeq
where $\tilde F^{l}$ is defined exactly as in \eqref{Fcech}
$\tilde F^{l} = dV^l - m^k_H C_2$ and the potential $V$ receives
two distinct contributions
\beq
V=
\tfrac{1}{2}\, 
(\text{Re} f)^{-1 kl} D_{k} D_{l} + 2\, m^k_H\text{Re} f_{kl}\, m^l_H\, L^2\ ,
\qquad
D_{k} = \big(e_k^H + 2\,\text{Im}f_{kl}\, m^l_H \big)\, L \ .
\eeq
The first term arises from eliminating the $D$-terms in 
the $U(1)$ field strength $W^k$ while the second term is a 
`direct' mass term for the scalar $L$.\footnote{Note that this second term
is a contribution to the potential which is neither a $D$- nor an
$F$-term but instead a `direct' mass term whose presence is enforced
by the massive two-form.}
Inserting the $D$-term yields a second contribution to the mass term
and one obtains altogether
\bea
V&=&
\frac12 \big[\big(e_k^H +2\text{Im} f_{kp}\, m^p \big)
(\text{Re} f)^{-1 kl} \big( e_l^H +2 \text{Im} f_{lr}\, m^r\big)
+4 m^k_H\, \text{Re} f_{kl}\, m^l_H\big] L^2\ . \qquad 
\eea 
Using \eqref{defLC}  and \eqref{gaugekin-O5} 
this precisely agrees with the second term in the potential \eqref{pot5}.

As before the first term in \eqref{pot5}  can be derived from  the superpotential
\eqref{W5}. Inserting \eqref{KpotL} into \eqref{Lsc} using \eqref{defLC}
 indeed yields the first term of  \eqref{pot5}.

\section{Conclusions}

In this paper we  determined the low energy effective action for Calabi-Yau 
orientifolds in the presence of background fluxes from a Kaluza-Klein reduction. 
In our analysis we did not specify a particular Calabi-Yau manifold but
merely demanded that it  admits  a holomorphic and isometric involution
$\sigma$.  Depending on 
the explicit action of $\sigma$ on the holomorphic three-form $\Omega$, 
we analyzed two distinct cases: (1) orientifolds with $O3/O7$-planes and (2) 
orientifolds with $O5/O9$-planes. 
For each case we calculated the K\"ahler potential, the 
superpotential and the gauge kinetic functions and showed the consistency
with $N=1$ supergravity.

In the first case the background fluxes induce a non-trivial scalar potential
which is determined in terms of a superpotential previously given in
\cite{GVW,TV,GKP,BBHL}.
We also included the scalar fields $(b^a,c^a)$ arising from the two
type IIB two-forms $B_2$ and $C_2$ and which are related to the presence of
$O7$-planes into the analysis. We showed that in this case the potential 
is unmodified which can be traced to the no-scale property of the K\"ahler potential.
This in turn (as well as the choice of the proper K\"ahler coordinates)
can be easily understood in terms of the `dual' formulation 
where the chiral multiplets containing the  K\"ahler deformations of the Calabi-Yau 
orientifold are replaced by linear multiplets. 
This formulation  of the effective action is particularly suitable for
also including the couplings of background $D$-branes to the bulk moduli
fields as given in \cite{DWG,Kachru:2003sx,GGJL}.
As a byproduct we determined an entire new class of no-scale
K\"ahler potentials which in the chiral formulation
can only be given implicitly as the solution of some constraint equation.
In the linear multiplet formalism on the other hand they are defined
straightforwardly and their no-scale property is easily displayed.

For orientifolds with $O5/O9$ planes the influence of background fluxes is more
involved. This is due to the fact that the space-time two-form $C_2$ arising in the 
expansion of the RR field $\hat C_2$ remains in the spectrum. It combines with 
the dilaton into a linear multiplet, 
which only if it is massless can be dualized to 
a chiral multiplet. However, generic NS three-form background fluxes render this
form massive. We therefore first restricted our attention to the case were the mass
term vanishes which occurs if the magnetic fluxes arising from
the NS three-form $H_3$ are set to zero. In the resulting
chiral description the axion dual to $C_2$ is gauged with the gauge charges set by the 
electric fluxes. The scalar potential now consists of two distinct contributions. 
The term which depends on the RR fluxes arising from 
$F_3$ is obtained from a (truncated) superpotential of the previous case
whereas the second contribution depends on the electric fluxes of
$H_3$ and arises from $D$-terms which are present  due to the gauged isometry.    
Finally, we also analyzed non-vanishing magnetic fluxes 
in the NS sector which can be described by an $N=1$ theory including a massive linear
multiplet coupled to vector and chiral multiplets. 
In this case the scalar potential as additional contributions
again arising from $D$-terms  but furthermore a direct mass term for the scalar  
in the linear multiplet which is neither a $D$- nor an $F$-term.

Orientifolds with $O3/O7$-planes can also be obtained as a limit of 
F-theory compactified on elliptic Calabi-Yau fourfolds \cite{Vafa,Sen}. 
At the level of the effective action this can indeed be seen by
compactifying the effective action given in section \ref{o3-action}
on a circle to $D=3$ and comparing it with the action
of M-theory compactified on Calabi-Yau fourfolds computed in 
\cite{HL}.
Notice that the K\"ahler potential in three dimensions
is the sum of the $D=4$ K\"ahler potential plus an additional term which
arises when the three-dimensional vectors are dualized to scalars.
Out of the vectors $V^\kappa$ present in the orientifold one
obtains complex scalars $w^\kappa$ in $D=3$, such that \cite{HL}
\beq\label{d3orienti}
  K_{(3)}=K_{(4)}(\tau,T,G,z) + K_{(vec)}(w,z)\ .
\eeq
Turning to the M-theory compactification, we recall that the massless spectrum 
of the fourfold $Y_4$ consists of $h^{(3,1)}$ complex structure deformations 
$Z$, $h^{(1,1)}$ complexified K\"ahler deformations $(M,P)$ and $h^{(2,1)}$ `non-geometrical'
scalars $N$. Together they span a K\"ahler manifold determined by a non-trivial
K\"ahler potential computed in \cite{HL}. In the orientifold limit this metric splits
into two parts and can be compared to the K\"ahler potential 
\eqref{d3orienti} of the compactified orientifold. 
One finds agreement if one identifies
\beq
Z \sim \tau,z^k\ , \qquad
(M,P)\sim T_\alpha \ , \qquad
N \sim G^a, w^\kappa \ .
\eeq
The inverse radius of the compactification circle can be
identified with the volume of the torus fiber. In lifting
the M-theory compactification to F-theory one shrinks the 
volume of the torus fiber \cite{Vafa}, which on the other
hand corresponds to going back to the four-dimensional orientifold. 
Further details about this F-theory lift will be presented elsewhere.


\vskip 1cm

\subsection*{Acknowledgments}

This work is supported by DFG -- The German Science Foundation,
the European RTN Programs HPRN-CT-2000-00148, HPRN-CT-2000-00122,
 HPRN-CT-2000-00131 and the
DAAD -- the German Academic Exchange Service.

We have greatly benefited from conversations and correspondence with 
S.J.\ Gates, M.\ Gra\~na, R.\ Grimm, H.\ Jockers, A.\ Micu and W.\ Schulgin. 
Moreover we like to thank Geert Smet and Joris Van Den Bergh for drawing our
attention to an error in an earlier version of this paper.

J.L.\ thanks E.\ Cremmer and the L.P.T.E.N.S.\ in Paris for hospitality
and financial support during the final stages of this work.
\\
\vspace{1cm}
\\
{\Large \bf Appendix}

\renewcommand{\theequation}{\Alph{section}.\arabic{equation}}
\appendix
\section{Conventions}\label{conventions}
In this appendix we summarize our conventions.

\begin{itemize}
\item
The coordinates of the four-dimensional Minkowski space-time are 
denoted by $x^\mu, \mu=0,\ldots,3$.
The corresponding metric is chosen to have signature $(-,+,+,+)$.
The coordinates of the compact Calabi-Yau manifold  $Y$
are $y^i, \bar y^\bi,\ i,\bi=1,2,3$. 

\item
$p$-forms are expanded into a real basis according to 
\bea
  A_p\ =\ \frac{1}{p!}\, 
A_{\mu_1 \ldots \mu_p} dx^{\mu_1}\wedge \ldots \wedge dx^{\mu_p}\ .
\eea
\item
$(p,q)$-forms are expanded into a complex basis as
\bea
  A_{p,q} = \frac{1}{p!q!} A_{i_1 \ldots i_p \bi_1 \ldots \bi_q} dy^{i_1}\wedge \ldots \wedge dy^{i_p}
            \wedge d\bar y^{\bi_1}\wedge \ldots \wedge d\bar y^{\bi_q}\ .
\eea
\item
The exterior derivative is defined as 
\bea
  dA_p=\frac{1}{p!} \partial_\mu A_{\mu_1 \ldots \mu_p} dx^\mu\wedge dx^{\mu_1}\wedge \ldots \wedge dx^{\mu_p}\ .
\eea
\item
The field strength of a $p$-form $F_{p+1}=dA_p$ 
is given by 
\bea
  F_{\mu_1 \ldots \mu_{p+1}} = (p+1)\, \partial_{[\mu_1}A_{\mu_2\ldots \mu_{p+1}]}\ .
\eea
\item
The inner product for real forms is defined by using the Hodge-$*$ operator. In
components we have 
\bea
 \int F_p \wedge * F_p = \frac{1}{p!}\int d^d x\, \sqrt{-g}\,  F_{\mu_1 \ldots \mu_p} F^{\mu_1 \ldots \mu_p}\ .
\eea
$*\mathbf{1} = d^d x\, \sqrt{-g}$ is the $d$-dimensional measure. 

\item
The Hodge-$*$ satisfies 
$** F_p = (-1)^{p(d-p)+\kappa} F_p$, where $\kappa=1$ for Lorentzian signature 
and $\kappa=0$ for Euclidean signature. 

\item 
On a Calabi-Yau manifold the non-trivial cohomology groups $H^{(p,q)}$ are
$$
H^{(0,0)},\ H^{(1,1)},\ H^{(3,0)},\ H^{(2,1)},\ H^{(1,2)},\ H^{(0,3)},\
 H^{(2,2)},\ H^{(3,3)}\ .
$$
Their dimensions $h^{(p,q)}$ obey
\beq
h^{(0,0)} =  h^{(3,0)}= h^{(0,3)}=h^{(3,3)}=1\ , \qquad
h^{(1,1)}=h^{(2,2)}\ , \qquad
h^{(2,1)}= h^{(1,2)}\ .
\eeq
For the indices labeling the basis elements of these cohomology groups  
we use the conventions
\bea
\omega_A \in H^{(1,1)}\  ,\qquad  \tilde \omega^A \in H^{(2,2)}\ ,\qquad  
A,B = 1, \ldots, h^{(1,1)} \  ,\nn\\
\chi_K \in  H^{(2,1)} \ , \qquad\bar \chi_K \in  H^{(1,2)} \ , \qquad
K,L = 1, \ldots, h^{(2,1)}\  , \\
(\alpha_{\hat K},\beta^{\hat L})\in   H^{(3)} \ , \qquad
\hat K, \hat L = 0, \ldots, h^{(2,1)}\  . \qquad  \qquad \nn
\eea

\item
For orientifolds the cohomology groups split according to
$H^{(p,q)}= H^{(p,q)}_+\oplus H^{(p,q)}_-$.
The basis elements are denoted as follows
\bea
\omega_\alpha \in H^{(1,1)}_+\  , &\quad  
\tilde \omega^\alpha \in H^{(2,2)}_+\ ,\qquad  
\alpha,\beta = 1, \ldots, h^{(1,1)}_+ \  ,\nn\\
\omega_a \in H^{(1,1)}_-\  , &\quad  
\tilde \omega^a \in H^{(2,2)}_-\ ,\qquad  
a,b = 1, \ldots, h^{(1,1)}_- \  ,\\
\chi_\kappa \in  H^{(2,1)}_+\  ,& 
\quad \kappa,\lambda = 1,\ldots,h^{(2,1)}_+\  ,\nn\\
\chi_k \in  H^{(2,1)}_-\  , & \quad 
k,l = 1,\ldots,h^{(2,1)}_-\  . \nn
\eea

For the cohomology $H^{(3)}$ there is a difference depending 
on the transformations property of $\Omega$.
For $\sigma^*\Omega=-\Omega$ ($O3/O7$ case) one has 
$\Omega \in H^{(3)}_-$ and thus $2 h^{(2,1)}_+ = h^{(3)}_+$
and $2h^{(2,1)}_- +2 = h^{(3)}_-$ holds.
As in the Calabi-Yau case we put a `hat' on the index if it
starts from $0$ and we have
\bea
(\alpha_\kappa,\beta^\lambda)\in
H^{(3)}_+, & \quad \kappa,\lambda = 1,\ldots,h^{(2,1)}_+\  ,\nn \\
(\alpha_{\hat k},\beta^{\hat l})\in   H^{(3)}_- \ , &\quad
\hat k,\hat l= 0,\ldots,h^{(2,1)}_- \  .
\eea

For $\sigma^*\Omega=\Omega$ ($O5/O9$ case) one has 
$\Omega \in H^{(3)}_+$ and thus $2 h^{(2,1)}_+ + 2 = h^{(3)}_+$
and $2h^{(2,1)}_- = h^{(3)}_-$ holds.
In this case we have 
\bea
(\alpha_{\hat\kappa},\beta^{\hat\lambda})\in
H^{(3)}_+, & \quad \hat\kappa, \hat\lambda = 0,\ldots,h^{(2,1)}_+\  ,\nn \\
(\alpha_{k},\beta^{l})\in   H^{(3)}_- \ , &\quad
\hat k,\hat l= 1,\ldots,h^{(2,1)}_- \  .
\eea

\end{itemize}

\bigskip

%
%

\section{The Potentials}\label{Spotentials}


\subsection{The scalar potential for the $O3/O7$ orientifolds} 
\label{scalarpot}

In this appendix we give some details about the derivation
of \eqref{potential37}. This already exists in the literature 
\cite{TV,GKP,BBHL,DWG} and we include it here in order to make the paper
more self-contained.

$V$ arises from the terms of \eqref{10d-lagr} given by 
\beq
  S^{(10)}=-\frac{1}{4} \int \left( e^{- \hat \phi} 
                  \hat H_3 \wedge *\; \hat H_3
                  +  e^{ \hat \phi} 
                  \hat F_3 \wedge *\; \hat F_3\right)+ \ldots\ .
\eeq
Inserting only the background fluxes $H_3$ and $F_3$
which are defined in \eqref{fluxes} and \eqref{HFF} and which are
harmonic three-forms on the Calabi-Yau manifold $Y$ with no 
four-dimensional space-time dependence we arrive at 
\beq\label{Vfirst}
 \cL = -\frac{1 }{4}\, e^{ \phi} \int_Y 
                 G_3 \wedge *_6\; \bar G_3  +\ldots \ ,
\eeq
where $*_6$ is the Hodge $*$-operator on $Y$.
One defines the imaginary self- and anti-self-dual parts of $G_3$ by
\begin{equation}\label{GSD}
  *_6 G^{\pm}_3 = \mp i G^{\pm}_3 \ , \qquad \text{i.e.}\ \ \ 
  G^{\pm}_3=\frac{1}{2}(G_3 \pm i *_6 G_3).
\end{equation}
Inserting \eqref{GSD} into \eqref{Vfirst} we arrive at
\beq\label{int}
\cL =     
    - \frac12\, e^{\phi}\int_{Y} G^+_3 \wedge *_6\; \overline{ G^+_3} 
  \ +  \frac{i}{4}\,  e^{\phi}\int_{Y} G_3 \wedge \bar G_3 + \ldots \ .
\end{equation}
The second term in \eqref{int} is a topological term 
contributing to the flux tadpoles while
the first term corresponds to the four-dimensional  scalar potential $V$.
Including the Weyl rescaling of the four-dimensional metric  
$g_{\mu \nu} \rightarrow \frac{\KK}{6} g_{\mu \nu}$
results in the overall factor $36/\KK^2$  and we obtain
\beq \label{S4_V}
V =  \frac{18}{\KK^2}\; e^{\phi}  
    \int_{Y} G^+_3 \wedge *_6\; \overline{G^+_3}\ .
\end{equation}

In eq.\ \eqref{fluxes} we observed that $G_3$ can be expanded
in a basis of  $H^{(3)}_-$.
The decomposition \eqref{GSD}
says that $G_3^+$ can be expanded along $H^{(3,0)}_-\oplus H^{(1,2)}_-$ 
and a choice of basis is $(\Omega, \bar \chi_{ k})$ where
the $\chi_{ k}$ are defined in (\ref{def-chi}).
Thus we have 
\begin{equation}
  G_3^+ = - \frac{1}{\int \Omega \wedge \bar \Omega} \left(
  \Omega \int \bar \Omega \wedge G_3 + G^{lk} \bar \chi_{k}
  \int \chi_{l} \wedge G_3
  \right), \label{G-expansion}
\end{equation}
where $G^{kl}$ is the inverse of the metric defined in (\ref{csmetrico})
and we have furthermore used 
$\int \bar \Omega \wedge G_3^+ = \int \bar \Omega \wedge G_3$ and
$\int \chi_{ k} \wedge G_3^+ = \int \chi_{ k} \wedge G_3$.
Inserting (\ref{G-expansion}) into (\ref{S4_V}) 
we finally arrive at the expression already given in \eqref{potential37}
\bea\label{pot-comp}
  V &=&\frac{18i\ e^{ \phi}}{\KK^2 \int \Omega \wedge \bar \Omega }
  \left( \int \Omega \wedge \bar G_3 \int \bar \Omega \wedge G_3 
  + G^{kl} \int \chi_{k} \wedge G_3 
  \int \bar \chi_{l} \wedge \bar G_3 \right) \ .
\eea

An alternative form of this potential can be obtained by 
inserting the explicit expansion of $G_3$ given in
\eqref{G3exp} into \eqref{Vfirst} and expressing the potential in terms of
the $2(h^{(1,2)}_-+1)$ complex flux parameters given in \eqref{mcomplex}.
Using \eqref{defperiod-m} and  \eqref{sbasiso} one finds \cite{TV}
\beq
V    = - \frac{9e^{\phi} }{\KK^2} 
  \Big[m^{\hat k} (\text{Im}\; \MM)_{\hat k \hat l} \, \bar m^{\hat l} +
  \big(e_{\hat k}- (m  \text{Re} \MM)_{\hat k}\big)   
\big(\text{Im} \MM\big)^{-1 \hat  k \hat l} 
  \big(\bar e_{\hat l}-(\bar m \text{Re} \MM )_{\hat l} \big) \Big]\ .
\eeq

\bigskip
\subsection{The scalar potential for $O5/O9$}
\label{Spotentials5}

Here we supply some details computing  the potential
\eqref{pot5}. The difference to the previous case is that
the two background fluxes $H_3$ and $F_3$ take values in different
cohomologies with no non-trivial intersections.
$F_3$  is expanded in a basis of $H^{(3)}_+$
while $H_3$ is expanded in a basis of $H^{(3)}_-$ 
given in \eqref{mef}. 
Since $\Omega$ is an element of $H^{(3)}_+$ 
the expansion of $F_3$ is completely analogous to 
the previous case  while $H_3$ is only expanded 
along $H^{(2,1)}_-$. As a consequence the derivation 
of the potential changes slightly.
First we decompose  $G_3=F_3-ie^{-\phi}H_3$ into
self-dual 
$G_{3}^+$ and anti self-dual $G_{3}^-$ components.
Analogous to \eqref{G-expansion} $G_{3}^+$ enjoys an expansion
\begin{eqnarray}\label{G-expansion-O5}
  G_{3}^+ &=&  F_3^+-ie^{-\phi}H_3^+\nn\\
&=&
- \frac{1}{\int \Omega \wedge \bar \Omega} \Big(
  \Omega \int \bar \Omega \wedge F_3 
+ G^{\lambda\kappa} \bar \chi_{\kappa}
  \int \chi_{\lambda} \wedge F_3 \Big)  \nn  \\
&&-\frac{i}{2} e^{- \phi} \Big( m^{k}_H \alpha_{k}-
  e_{k}^H \beta^{k} + 
  i (m^{k}_H *\alpha_{k}-e_{k}^H *\beta^{k})\Big)\\
&=& -\frac{1}{2}  \Big( m^{\hat\kappa}_F \alpha_{\hat\kappa}-
  e_{\hat\kappa}^F \beta^{\hat\kappa} + 
  i (m^{\hat\kappa}_F *\alpha_{\hat\kappa}
-e_{\hat\kappa}^F *\beta^{\hat\kappa})\Big)  \nn\\
&&-\frac{i}{2} e^{- \phi} \Big( m^{k}_H \alpha_{k}-
  e_{k}^H \beta^{k} + 
  i (m^{k}_H *\alpha_{k}-e_{k}^H *\beta^{k})\Big) \ ,\nn
\end{eqnarray}
where we have used \eqref{G-expansion} and 
the self-dual combinations of \eqref{mef}.
Inserted into \eqref{S4_V} and using the fact that 
$H_3$ and $F_3$ have no non-trivial intersections we arrive at
\bea
V &=&
  \frac{18i\ e^{\phi}}{\KK^2 \int \Omega \wedge \bar \Omega }
  \left( \int \Omega \wedge F_3 \int \bar \Omega \wedge F_3 
  + G^{\kappa \lambda} \int \chi_{\kappa} \wedge F_3 
  \int \bar \chi_{\lambda} \wedge F_3 \right) \\
  && 
  - \frac{9e^{-\phi}}{\KK^2} \Big[
  m^{k}_H (\text{Im} \MM)_{ k l} \,  m^{l}_H +
  \big(e_{k}^H-(m_H \text{Re} \MM)_{k}\big)   
\big(\text{Im} \MM\big)^{-1  k l} 
  \big(e_{l}^H-(m_H \text{Re} \MM)_{l}\big) \Big] \nn \\
 &=&
  - \frac{9e^{\phi}}{\KK^2} \Big[
  m^{\hat \kappa}_F (\text{Im}\; \MM)_{ \hat \kappa \hat \lambda} \,  
m^{\hat \lambda}_F +
  \big(e_{\hat \kappa}^F-(m_F \text{Re} \MM)_{\hat \kappa}\big)   
  \big(\text{Im} \MM\big)^{-1 \hat \kappa  \hat \lambda} 
  \big(e_{\hat \lambda}^F-(m_F \text{Re} \MM)_{\hat \lambda}\big) \Big]\nn\\ 
 && - \frac{9e^{-\phi}}{\KK^2} \Big[
  m_H^{k} (\text{Im} \MM)_{ k l} \,  m_H^{l} +
  \big(e^H_{k}-(m_H \text{Re} \MM)_{k}\big)   
\big(\text{Im} \MM\big)^{-1  k l} 
  \big(e^H_{l}-(m_H \text{Re} \MM)_{l}\big) \Big] \ .\nn 
\eea

%
%

\section{The K\"ahler metrics}\label{Kmetrics}

\subsection{The K\"ahler metric and its inverse for $O3/O7$}
\label{Kmetrics3}
In this appendix we supply the details for the the calculation of the
K\"ahler metric obtained from the K\"ahler potential $K(\tau,T,G,z)$ given in 
\eqref{kaehlerpot-O7-1}. 
It turns out to be instructive to do this computation for a more general 
K\"ahler potential than the one given in \eqref{kaehlerpot-O7-1}
and only in the end restrict ourselves to this case.

Let $M^I=(T_\alpha, N^A)$ be the coordinates 
of a K\"ahler manifold with  K\"ahler potential 
\beq\label{KPo}
 K  =  K_0(N, \bar N) + K_\KK(T+\bar T, N, \bar N) \ ,
\eeq
where $K_\KK$ has the special form
\beq
 K_\KK = - 2 \text{ln}\big(\cK\big)= 
- 2 \text{ln}\big(\cK_{\alpha \beta \gamma} v^\alpha v^\beta v^\gamma\big) \ .
\label{KPo2}
\eeq
The $v^\alpha(T+\bar T,N,\bar N)$  are functions of the K\"ahler coordinates
 defined implicitly by the 
equation 
\bea
 T_\alpha =  \frac{3i}{2} \rho_\alpha + \frac{3}{4} \KK_{\alpha \beta \gamma} v^\beta v^\gamma - \frac{3}{2} \zeta_\alpha(N,\bar N)\ ,
  \label{eT_alpha}
\eea
where the $\zeta_\alpha$ are arbitrary 
complex functions depending on the coordinates $N^A,\bar N^A$
but not on $T_\alpha$. 
We see that for 
\beq\label{O3-case}
K_0 = K_{\rm cs} -\ln\big[-i(\tau-\bar\tau)\big]\ ,\qquad
\zeta_\alpha = 
   -\frac{i}{2(\tau-\bar\tau)} \KK_{\alpha bc} G^b(G-\bar G)^c
\eeq
the $K$ of \eqref{KPo} coincides with the $K$ given in 
\eqref{kaehlerpot-O7-1}.

The K\"ahler metric is computed by differentiating  the real part 
of \eqref{eT_alpha} with respect to all K\"ahler coordinates $T_\alpha,N^A$.
This yields
\bea
  \frac{\partial v^\alpha}{\partial T_\beta} = \frac{1}{3} \KK^{\alpha \beta} \ ,
  \qquad
  \frac{\partial v^\alpha}{\partial N^A} = \frac{1}{2}\, \KK^{\alpha \beta} 
  \zeta_{\beta,A}^R\ , \label{dv_dN}
\eea
where  $\zeta^R_\alpha \equiv \zeta_\alpha + \bar \zeta_\alpha$
and $ \KK^{\alpha \beta} $ is the inverse of 
$ \KK_{\alpha \beta} =\KK_{\alpha \beta\gamma}v^\gamma$. 
Using  (\ref{dv_dN}) one computes the first derivatives 
of \eqref{KPo} to be
\bea \label{K-deriv}
  K_{T_\alpha}=-2\, \frac{v^\alpha}{\cK}\ , \qquad  K_{N^A}=K_{0, A}- 3\, \frac{v^\alpha}{\cK}\, \zeta^R_{\alpha,A}\ .
\eea
Similarly, the K\"ahler metric is found to be
\bea
  K_{T_\alpha \bar T_\beta} &=& \frac{G^{\alpha \beta}}{\KK^2}\ , \quad
  K_{T_\alpha \bar N^A} \ = \ \frac{3}{2}\frac{G^{\alpha \beta}}{\KK^2}\, \zeta^R_{\beta,\bar A}\ , \nn \\
  K_{N^A \bar N^B} &=& K_{0 ,A \bar B} + Q_{A\bar B}
  + \frac{9}{4} \frac{G^{\alpha \beta}}{\KK^2} \zeta^R_{\alpha,A}\, \zeta^R_{\beta,\bar B}\ ,\label{KMet}
\eea
where $G^{\alpha \beta}$ is given in \eqref{Ginvers} 
and we abreviated 
$Q_{A\bar B}=-3 \frac{v^\alpha}{\cK} \zeta^R_{\alpha,A \bar B}$.
Finally, using  (\ref{KMet}) and (\ref{eT_alpha}) one can express
the K\"ahler metric in the (non-K\"ahler) coordinates
$(v^\alpha,\rho_\alpha, N^A)$ 
\bea \label{scalaract}
&&  K_{M^I \bar M^J} dM^I \wedge * d\bar M^J   =\\
 &&\qquad \qquad  G_{\alpha \beta}\, dv^\alpha \wedge * dv^\beta 
   +\frac{1}{4} d\, \text{ln}\KK \wedge * d\, \text{ln} \KK  
+\big(K_{0 ,A \bar B}
    - 3\frac{v^\alpha}{\KK} \zeta^R_{\alpha,A \bar B}\big)\, dN^A \wedge * d \bar N^B  \nn \\ 
&& \qquad\qquad   +\frac{9}{4} \frac{G^{\alpha \beta}}{\KK^2} \Big(d\rho_\alpha - 
   i(\bar \zeta_{\alpha, A}dN^A - \zeta_{\alpha, \bar A} d\bar N^A)\Big)\wedge 
   * \Big(d\rho_\beta - 
   i(\bar \zeta_{\beta, B}dN^B - \zeta_{\beta, \bar B} d\bar N^B)\Big). \nn
\eea

Let us now discauss two special cases. First we assume that the K\"ahler 
manifold is a direct product with block-diagonal metric and $K_0$ ($K_\KK$) is
the K\"ahler potential of the first (second) factor. An example for this situation 
is the case $\tau=\text{const}$.
In this case (for arbitrary $\zeta_{\alpha}$) the K\"ahler metric of
the K\"ahler potential $K_\KK$  can be inverted as
\bea \label{invME}
  K_\KK^{T_\alpha \bar T_\beta} = \KK^2 G_{\alpha \beta} + \frac94 
                              Q^{A \bar B} \zeta^R_{\alpha,A} \zeta^R_{\beta,\bar B}\ , \quad
  K_\KK^{T_\alpha \bar N^A} = - \frac32 Q^{\bar A B} \zeta^R_{\alpha,B}\ , \quad
  K_\KK^{N^A \bar N^B} = Q^{A \bar B}\ ,
\eea
where $Q^{A \bar B}$ is the inverse of $Q_{A \bar B}$. 
Using \eqref{K-deriv} and \eqref{invME} one finds
\bea
  \frac{\partial K_\KK}{\partial M^I}\ (K^{-1})_\KK^{I \bar J}\ \frac{\partial K_\KK}{\partial \bar M^{J}}\ =\ 3\ .
\eea 
This implies that $K_\KK$ obeys the standard no-scale condition \cite{NS}.
 
The second case corresponds to a K\"ahler potential given in \eqref{kaehlerpot-O7-1}.
One inserts \eqref{O3-case} and  $N^A=(\tau, G^a, z^k)$
into \eqref{KMet} to obtain 
\bea\label{KmetricO7}
   K_{T_\alpha \bar T_\beta} &=& \tfrac{1}{\cK^2} G^{\alpha \beta}\ ,\nn \\
   K_{T_\alpha \bar G^a} &=& -\tfrac{3i}{2\cK^2} 
                              G^{\alpha \beta} \cK_{\beta a b} b^b\ , \nn\\
   K_{T_\alpha \bar \tau} &=& -\tfrac{3i}{4\cK^2} G^{\alpha \beta} 
                              \cK_{\beta a b} b^a b^b\ , \nn\\
   K_{G^a \bar G^b} &=& e^{\phi} G_{ab}+\tfrac{9}{4\cK^2} 
                        G^{\alpha \beta} \cK_{\alpha a c} b^c
                        \cK_{\beta b d} b^d\ , \\
   K_{G^a \bar \tau} &=& e^{\phi} G_{ab}b^b+\tfrac{9}{8\cK^2} 
                         G^{\alpha \beta} \cK_{\alpha a c} b^c
                         \cK_{\beta b d} b^b b^d\ , \nn\\
   K_{\tau \bar \tau} &=& \tfrac{1}{4} e^{2 \phi} +
                          e^{ \phi} G_{ab}b^a b^b +\tfrac{9}{16\cK^2} G^{\alpha \beta}
                          \cK_{\alpha a c} b^a b^c
                          \cK_{\beta b d} b^b b^d\ , \nn\\
   K_{z^{k} \bar z^{l}} &=& G_{kl}\ ,\nn
\eea
and the inverse metric
\bea\label{Kinvers}
   K^{T_\alpha \bar T_\beta} &=& \cK^2 G_{\alpha \beta} + \tfrac{9}{4}
  e^{- \phi} G^{ab} \cK_{\alpha a c} b^c
   \cK_{\beta b d} b^d + \tfrac{9}{4} e^{-2 \phi} \cK_{\alpha a c} b^a b^c
   \cK_{\beta b d} b^b b^d\ , \nn\\
   K^{T_\alpha \bar G^a}&=&-\tfrac{3i}{2} e^{- \phi} 
  G^{ab} \cK_{\alpha b c}b^c - 3i e^{-2 \phi}\cK_{\alpha b c} b^b b^c b^a\ , \nn \\
   K^{T_\alpha \bar \tau}&=&3i e^{-2 \phi} \cK_{\alpha b c}b^c \ , \nn \\
   K^{G^a \bar G^b} & = &e^{-\phi} G^{ab} + 4 e^{-2\phi} b^a b^b\ ,\nn \\
   K^{G^a \bar \tau} & = &-4 e^{-2 \phi} b^a \\
   K^{\tau \bar \tau} & = & 4e^{-2\phi}\ , \nn \\
   K^{z^{k}\bar z^{l}} &=&G^{kl} \nn\ .
\eea
Using \eqref{Kinvers} and \eqref{K-deriv} one verifies \eqref{NScond}.
In section \ref{Linearm} we trace this property to
the `dual' formulation where instead of 
the chiral superfields $T_\alpha$ one uses the dual linear multiplets 
$L^\alpha$.


\subsection{The K\"ahler metric and its inverse for $O5/O9$}
\label{Kmetrics5}
For completeness let us also give explicitly the K\"ahler metric
of the K\"ahler potential 
\eqref{O5-Kaehlerpot} expressed in the (non-K\"ahler) variables 
$(v^\alpha,c^\alpha, \rho_a, b^a, \phi,h,z^\kappa)$.
By straightforward differentiation one finds for the metric
\bea\label{KmetricO5}
 K_{t^\alpha \bar t^\beta} &=& e^{\phi} G_{\alpha \beta}+\tfrac{9}{4\cK^2}
 G^{ab}\cK_{\alpha ac}b^c \cK_{\beta bd}b^d
 + \tfrac{9}{4\cK^2}e^{-\phi}\cK_{\alpha ab} b^a b^b \cK_{\beta cd} b^c b^d
 \ ,\nn\\
  K_{A_a \bar t^\alpha} &=& -\tfrac{9}{2\cK^2}
     e^{-\phi}\cK_{\alpha cd}b^c b^d b^a
  -\tfrac{9}{4\cK^2} G^{ac}\cK_{\alpha cd}b^d\ , \nn\\
  K_{S \bar t^\alpha} &=& -\tfrac{9}{2\cK^2}
                           e^{-\phi}\cK_{\alpha ab}b^a b^b\ , \nn\\
   K_{A_a \bar A_b} &=& \tfrac{9}{4\cK^2} G^{ab}+
                            \tfrac{9}{\cK^2} e^{- \phi} b^a b^b\ , \\
   K_{A_a \bar S} &=& \tfrac{9}{\cK^2} e^{- \phi}b^a\ , \nn\\
   K_{S \bar S} &=&  \tfrac{9}{\cK^2} e^{- \phi}\ , \nn\\
   K_{z^{\kappa} \bar z^{\lambda}} &=& G_{\kappa \lambda}\ .\nn
\eea
Inverting this metric yields 
\bea \label{KinvO5}
  K^{t^\alpha \bar t^\beta} &=& e^{-\phi} G^{\alpha \beta}\ ,\nn\\
  K^{t^\alpha \bar A_a} &=& e^{-\phi} G^{\alpha \beta} \cK_{\beta a b}b^b, \nn\\
  K^{t^\alpha \bar S} &=& -\tfrac{1}{2} e^{-\phi} G^{\alpha \beta} \cK_{\beta ab}b^a b^b, \nn\\
  K^{A_a \bar A_b} &=& \tfrac49  \KK^2 G_{ab} 
                     + e^{-\phi} G^{\alpha \beta} \KK_{\alpha a c} b^c \KK_{\beta b d} b^d\ , \\
  K^{A_a \bar S} &=& - \tfrac49 \KK^2 G_{ab} b^b 
                     -\tfrac12 e^{-\phi} G^{\alpha \beta} \KK_{\alpha a c} b^c \KK_{\beta b d}b^b b^d\ , \nn\\
  K^{S \bar S} &=& \tfrac{1}{9} e^\phi \KK^2  +  \tfrac49  \KK^2 G_{ab} b^a b^b 
                   + \tfrac14  e^{-\phi} G^{\alpha \beta} \KK_{\alpha a c}b^a b^c \KK_{\beta b d}b^a b^d\ , \nn\\
  K^{z^{\kappa} \bar z^{\lambda}} &=& G^{\kappa \lambda}\ .\nn
\eea
Notice that the $O5/O9$ metric for $(S, t^\alpha, A_a)$ in non-K\"ahler coordinates is
up to factors $\KK$ and $e^\phi$ the inverse of the $O3/O7$ metric for $(\tau, T_\alpha, G^a)$.
This can be understood by noticing that $T_\alpha$ and $t^\alpha$ 
are 'dual' coordinates \cite{DAFT}.


\end{document}